\documentclass[useAMS,usenatbib]{mn2e_mod}
\usepackage{natbib}
\usepackage{graphicx}
\usepackage{amsmath}
\usepackage{amssymb}
\usepackage{deluxetable}
\usepackage{longtable}
\usepackage{url}

\let\ACMmaketitle=\maketitle
\renewcommand{\maketitle}{\begingroup\let\footnote=\thanks \ACMmaketitle\endgroup}

\title [IR properties of winds]{Discovering AGN-driven winds through their infrared emission: I. General method and wind location}

\author[Baron \& Netzer]
{Dalya Baron$^{1}$\thanks{dalyabaron@gmail.com} \&
Hagai Netzer$^{1}$
\\
\\
$^{1}$School of Physics and Astronomy, Tel-Aviv University, Tel Aviv 69978, Israel.\\
}

\begin{document}

\maketitle

\label{firstpage}
\begin{abstract}

Large scale outflows of different gas phases are ubiquitous in the host galaxies of active galactic nuclei (AGN). Despite their many differences, they share a common property - they all contain dust. The dust is carried with the outflow, heated by the AGN, and emits at infrared wavelengths. This paper shows that the infrared emission of this dust can be used to detect AGN outflows and derive their properties. We use a sample of $\sim$4\,000 type II AGN and compare the infrared properties of systems that show spectroscopic signature of ionized gas outflows to systems that do not. We detect an additional mid-infrared emission component in galaxies with spectroscopically discovered winds, and attribute it to the dust in the outflow. This new component offers novel constraints on the outflow properties, such as its mean location and covering factor. We measure the location of the outflow for $\sim$1\,700 systems, with the distribution showing a prominent peak around $r \sim 500$ pc, a tail that extends to large distances ($\sim$10 kpc), and no objects with a location smaller than 50 pc. The covering factor of the wind shows a wide distribution which is centered around 0.1, with 24\% (8\%) of the winds showing covering factors larger than 0.2 (0.5). The dust emission is not sensitive to various systematics affecting optically-selected outflows, and can be used to estimate the mass outflow rate in thousands of galaxies with only 1D spectra. 

\end{abstract}


\vspace{1cm}
\section{Introduction}\label{s:intro}

AGN-driven winds are ubiquitous. They are routinely detected in host galaxies of active galactic nuclei (AGN) of all types and luminosities (e.g., \citealt{greene05, nesvadba06, mullaney13, rupke13, cicone14, cheung16, zakamska16, fiore17, rupke17, baron17b, baron18}). They appear in different gas phases; molecular, atomic and ionized gas (e.g., \citealt{greene11, cano12, veilleux13, harrison14, rupke17}) and have been invoked in the context of co-evolution of super massive black holes (SMBH) and their host galaxies as a way to couple the energy output of an accreting SMBH to the interstellar medium of the host galaxy (e.g., \citealt{silk98, kauffmann00, zubovas14}), thus explaining the observed correlations between the masses of SMBHs and their hosts (e.g., \citealt{magorrian98, ferrarese00, gebhardt00a, tremaine02}; for a recent review see \citealt{kormendy13}).

In order to assess the effect of such flows on their host galaxy evolution, it is necessary to determine their occurrence rate, disentangle AGN-driven and star formation-driven winds, and collect information about the outflow velocities, covering factors, mass outflow rates, and kinetic powers. The largest samples of galactic-scale winds in active galaxies come from spatially-integrated (1D) spectroscopy (e.g., \citealt{mullaney13, harrison16, woo16}). Such data can constrain the prevalence of winds in AGN host galaxies, and statistically determine the relative contribution of the AGN and the star formation to the outflow. However, it is usually impossible to use 1D spectroscopy to constrain the mass outflow rate and the kinetic power of these winds. 

Spatially-resolved 2D spectroscopic observations of AGN-driven winds, either by long slit observations or with integral field units (IFUs), have been used to overcome most of these difficulties \citep{fischer11, liu13a, liu13b, harrison14, karouzos16a, karouzos16b, fiore17, baron18}. 2D observations are expensive and todays long slit and IFU-based samples are significantly smaller than the 1D spectroscopic samples. Furthermore, IFU-based observations suffer from various uncertainties and systematics. In particular, the observed velocity and size of the wind are subjected to projection effects and outflows with small inclination angles are easier to detect and measure, because of the larger line-of-sight velocity (e.g., \citealt{harrison14, karouzos16a}). In addition, dust extinction can significantly affect the receding part of the wind  making it difficult to estimate the total mass outflow rate \citep{karouzos16a, baron18}.

The nature of multi-phased (molecular, atomic and ionized gas) outflows is largely unconstrained (e.g., \citealt{cicone18}). It is unclear whether these phases are physically connected, and different phases can have different outflow velocities, covering factors, and mass (e.g., \citealt{fiore17}). The one thing they have in common is that they all contain dust (e.g., \citealt{karouzos16a, baron17b, baron18, mccormick18}). This dust is carried by the outflow, heated by the AGN radiation, and must emit at infrared wavelengths. Such an emission is unavoidable, and the only question is its detectability. This leads to a relatively simple and novel method, the center of this paper, to detect AGN-driven winds by their dust emission.

The infrared emission of the outflowing dust is not subject to projection effects, allowing the detection of outflows that are perpendicular to the line of sight. It can be used to constrain the outflow location and covering factor, and since infrared photons are hardly affected by extinction, one can trace the full extent and mass of the wind. Various studies examined the colder dust component that is associated with galactic-scale winds in non-AGN systems using FIR observations (e.g., \citealt{tacconi_garman05, roussel10, mccormick13, melendez15, mccormick18}). In AGN, outflows were examined in the context of global obscuration (obscured versus unobscured AGN) via MIR selection \citep{dipompeo18}. To the best of our knowledge, an extensive study of AGN outflows, based on their dust emission properties, has never been done. 

In this work we study the infrared properties of AGN-driven winds, with the goal of characterizing the dust and ionized gas properties. We make use of publicly-available catalogs of $\sim$4\,000 type II AGN, and compare the infrared spectral energy distributions (SEDs) of systems which show AGN-driven winds to those that do not. We describe our sample in section \ref{s:sample}, compare windy and non-windy type II AGN in section \ref{s:infrared_sed}, and investigate the wind properties in section \ref{s:wind_properties}. We discuss our results in the context of general AGN-driven winds in section \ref{s:disc}, and conclude in section \ref{s:concs}.

Throughout this work we make extensive use of the term "covering factor" which we relate to various components. We refer to it as a simple way to measure the IR energy emitted by the dust in this component relative to the total energy output of the AGN. This can be very different, and smaller, from the geometrical covering factor of this component, depending on the dust optical depth and the fraction of the radiation absorbed by the gas. Thus, the covering factors of the central torus, the ionized outflowing gas and the NLR might not indicate the same physical property and should not be compared without taking into account the above additional factors. 

\begin{figure}
\includegraphics[width=3.25in]{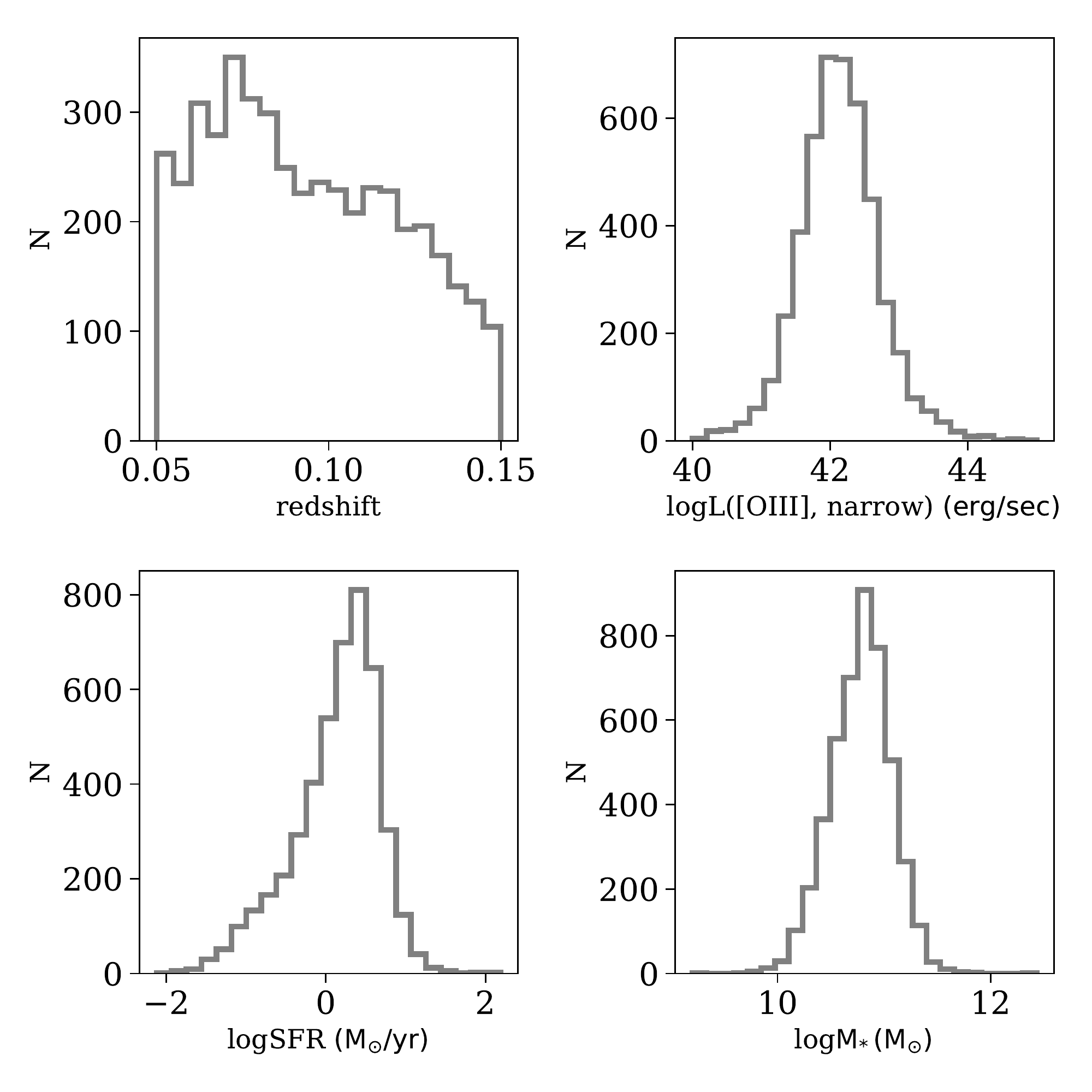}
\caption{The properties of the initial sample of type II AGN before division into windy and non-windy sub-groups. Redshifts (upper left) and reddening corrected narrow L([OIII]) luminosity (upper right) are both taken from the ALPAKA catalog. The SFR (bottom left) and the stellar mass (bottom right) are taken from the MPA-JHU catalog.}\label{f:initial_sample_histograms}
\end{figure}

\section{Sample Selection}\label{s:sample}

Our goal is to compare the infrared properties of type II AGN that show wind signature in their optical spectra to type II AGN that do not. We start with the publicly available catalog: AGN Line Profile And Kinematics Archive (ALPAKA\footnote{\url{https://sites.google.com/site/sdssalpaka/home}}; \citealt{mullaney13}), which provides emission line measurements for a sample of 24\,264 optically-selected AGN. The catalog uses all the extragalactic spectra from SDSS DR7 \citep{abazajian09} for which the optical emission lines [OIII]$\lambda$5007\AA, [NII]$\lambda$6584\AA, H$\alpha$, and H$\beta$ were detected to at least 3$\sigma$ by the automatic fitting procedure of the SDSS. For each spectrum, they fit the local continuum emission with a 5 degree polynomial and subtract it to obtain the emission line spectrum. Next, they perform a multi-component fitting to the emission line spectra which includes both narrow and broad kinematic components, and re-measure the intensity and the profile of [OIII]$\lambda$5007\AA, [NII]$\lambda$6584\AA, H$\alpha$, and H$\beta$. The catalog provides classifications of type I and type II AGN for the entire sample (see additional details in \citealt{mullaney13}). In this work we use the luminosities of the broad and narrow emission lines provided by the catalog.

We select systems which are classified by the ALPAKA catalog as type II AGN in the redshift range: $0.05 \leq z \leq 0.15$. The lower limit was chosen to ensure that the SDSS 3 arc-sec fiber covers most of the galaxy, so that systems with outflow will show wind signatures in their spectrum. The upper limit was chosen in order to avoid selection effects since many type II AGN drop from the SDSS sample beyond a redshift of 0.15. We further select systems in which the narrow [OIII] luminosity is measured to more than 3$\sigma$ by the ALPAKA catalog (there is some disagreement between the SDSS and the ALPAKA catalog for a small number of objects). 

We cross-match this sample with the MPA-JHU catalog, which provides stellar mass and star formation rate (SFR) measurements of SDSS DR7 galaxies \citep{b04, kauff03b, t04, salim07}. The stellar mass was measured using both stellar population synthesis fits to the galaxy spectra and broad band photometry (see \citealt{kauff03b} for details). The SFR was measured using a combination of the stellar mass and the Dn4000 index \citep{b04}, and an improved aperture corrections to account for the light outside the SDSS fiber \citep{salim07}. We use the 16th, 50th, and 84th percentiles of the stellar mass and SFR estimates provided by the MPA-JHU catalog.

\begin{figure}
\includegraphics[width=3.25in]{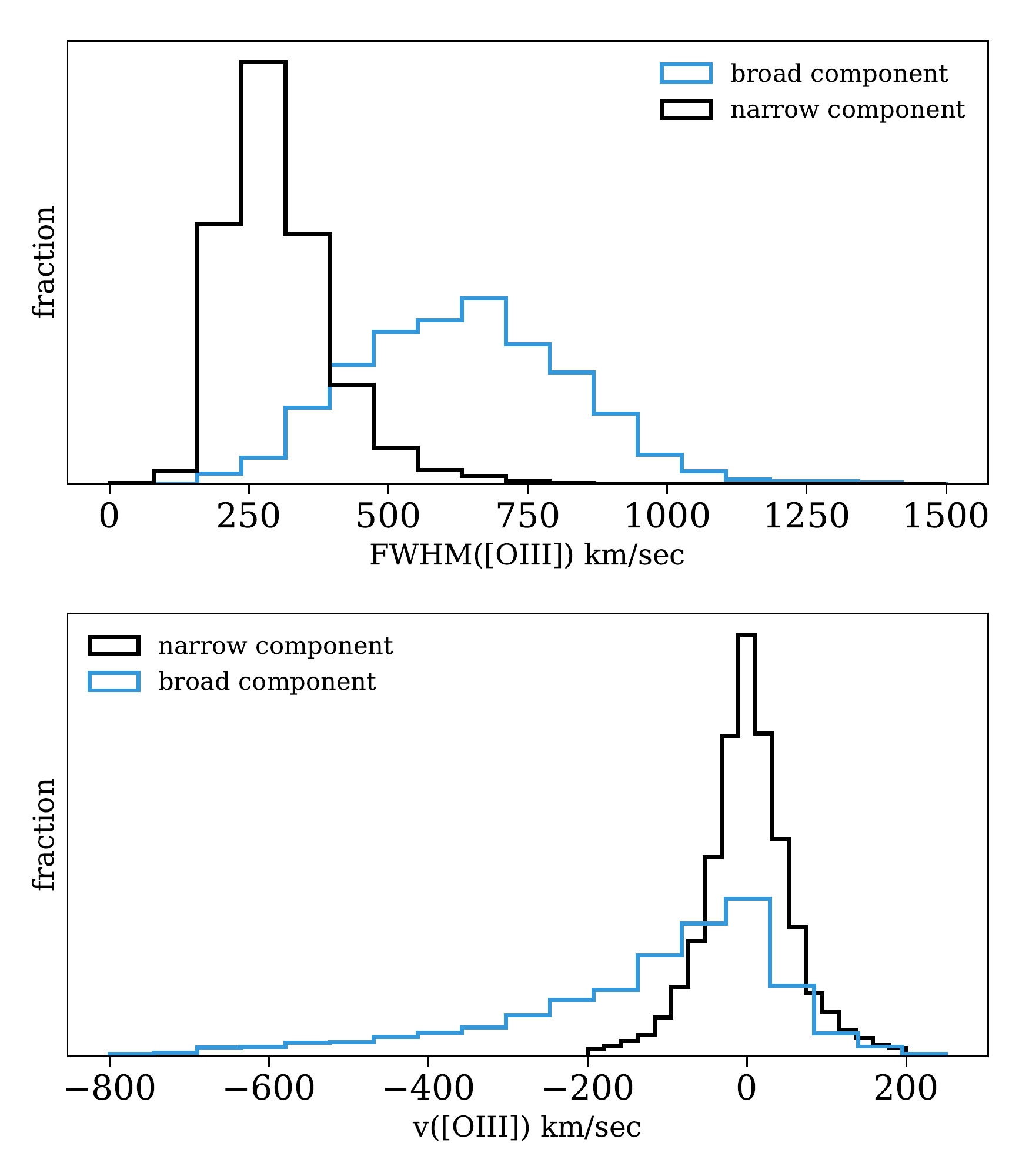}
\caption{Top panel: the distribution of FWHM([OIII]) for the narrow (black histogram) and broad (blue histogram) components in our sample. Bottom panel: the distribution of the velocity shift of the line centroid with respect to the systemic redshift of the system in our sample. We refer the reader to \citealt{mullaney13} for additional details regarding the multicomponent fitting process.}\label{f:oiii_vel_distribution_broad_and_narrow}
\end{figure}

Our final sample consists of 4\,580 type II AGN, for which we have the following measurements: narrow and broad [OIII]$\lambda$5007\AA\, luminosity, narrow H$\alpha$ and H$\beta$ luminosity, SFR, stellar mass, and broad band photometry. We show in figure \ref{f:initial_sample_histograms} histograms of the redshift, dust-corrected narrow [OIII] luminosity, stellar mass, and SFR of this sample. We divide the sample into two groups, which we call \emph{non-windy} and \emph{windy}. The windy galaxies show evidence of an outflow in their [OIII] line, as indicated by additional broader components to the line profile. The non-windy galaxies show no such components, and are well fitted with a single Gaussian profile. In the top panel of figure \ref{f:oiii_vel_distribution_broad_and_narrow} we show the distribution of the FWHM of the [OIII] emission line for the two components. The bottom panel shows the velocity shift of the components with respect to the systemic redshift of the system. We refer the reader to \citealt{mullaney13} for details regarding the multicomponent fitting procedure. 

We find 2\,203 type II AGN which do not show winds and 2\,377 which do. Due to the difficulty in detecting a broad [OIII] component in the SDSS spectra, we also experimented with a division into three groups: \emph{non-windy}, where no broad component is detected, \emph{weak winds}, where the broad component is detected but its flux is lower than the flux of the narrow component, and \emph{strong winds}, where a broad component is detected and its flux exceeds that of the narrow component. We found no difference between the \emph{weak winds} and the \emph{strong winds} groups, thus we focus only on the windy and non-windy groups. 

\section{Infrared SED}\label{s:infrared_sed}

We make use of optical and infrared (IR) photometric data as follows. We use the $riz$ optical photometry by the SDSS, the $JHK$ NIR photometry from the 2-Micron All-Sky Survey (2MASS; \citealt{skrutskie06}), and the W1-W4 MIR photometry from WISE \citep{wright10}. The effective wavelengths of the 2MASS bands are 1.22, 1.63, and 2.19 $\mathrm{\mu m}$, and the effective wavelengths of the WISE bands are 3.368, 4.618, 12.082, and 22.194 $\mathrm{\mu m}$. The SDSS provides astrometric cross-matches\footnote{\url{https://skyserver.sdss.org/CasJobs}} of objects observed by the SDSS and by the 2MASS and WISE surveys. Specifically, we extract the photometry from the 2MASS All-Sky Extended Source Catalog, since most galaxies are spatially resolved, and the WISE All-Sky Source Catalog. Out of the initial sample, 3.5\% of the objects do not have an astrometric cross-match by the SDSS. Out of the remaining, 15\% of the objects are not detected in the W4 WISE band, and 1\% of the objects are not detected in both W3 and W4 bands. We discgard systems which are not detected in their W3 and W4 bands, but keep systems with a non-detection in the W4 band. For the latter, we consider only their $riz$, $JHK$, and W1-W3 bands. We converted the photometric measurements and uncertainties into fluxes and propagated the uncertainties accordingly. We then used the measured redshifts to obtain $\lambda \mathrm{L}_{\lambda}$ in the centers of each of the bands. We do not correct for Galactic dust reddening since most objects are at a high galactic latitude, where dust reddening is negligible, and since the infrared bands suffer very little extinction.

The analyzed SEDs are a combination of several different components. At optical and NIR wavelengths, the SEDs are dominated by the direct stellar light, which is connected to the stellar mass of the system. At intermediate wavelengths (MIR), the SED is dominated by the torus emission, which is connected to the bolometric luminosity of the AGN and also to the [OIII] luminosity. This band also contains contributions from NLR dust and a new dust component, introduced here for the first time, due to the outflow. At longer wavelengths (MIR and FIR), the SED is dominated by dust that is heated by O- and B-type stars, and is related to the SFR. 

\begin{figure}
\includegraphics[width=3.25in]{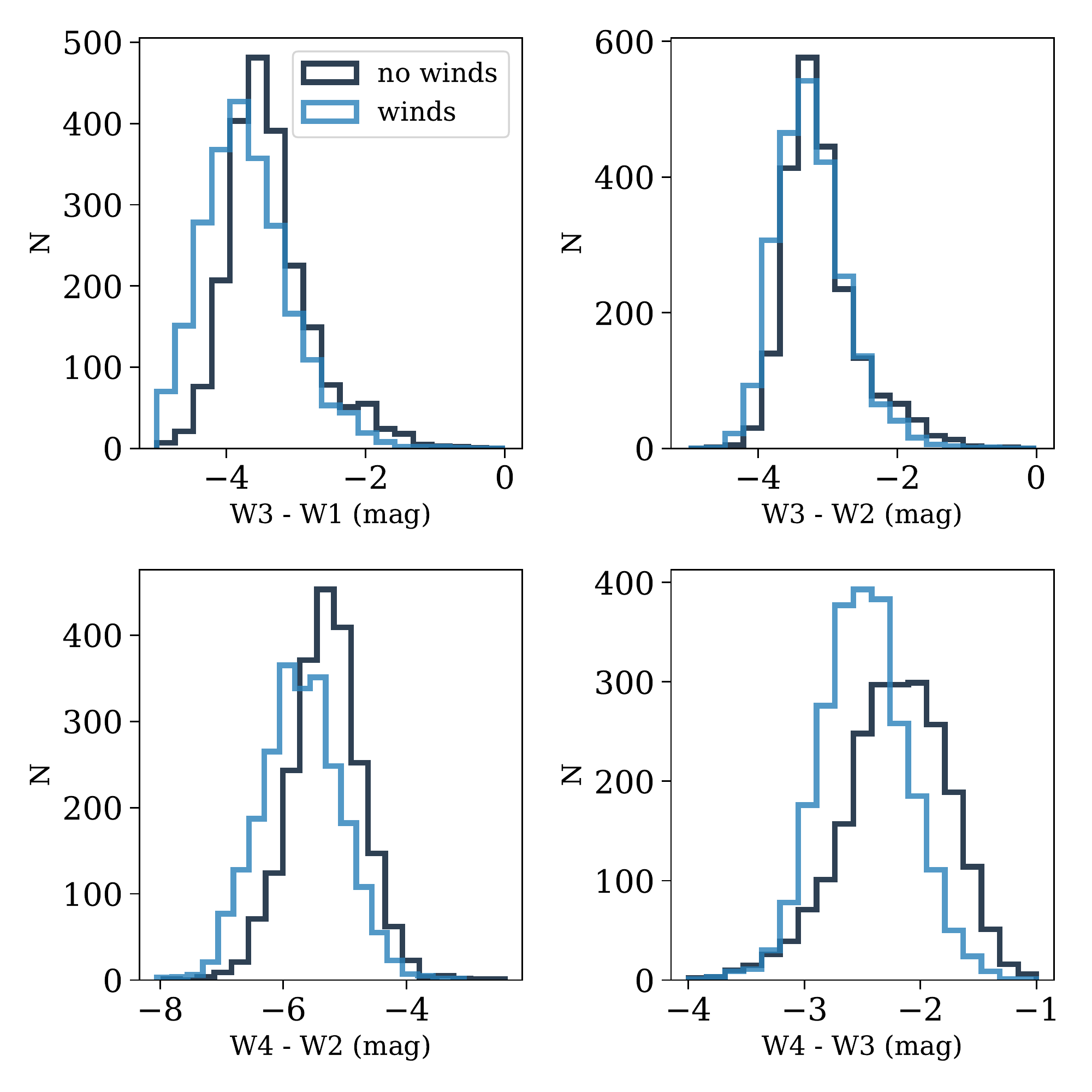}
\caption{Comparison of the WISE colors of the windy and non-windy type II AGN. In the upper panel we show the distribution of W3-W1 (left) and W3-W2 (right), and in the lower panel the distribution of W4-W2 (left) and W4-W3 (right). Windy and non-windy systems show slightly different MIR colors, except for W3-W2.}\label{f:wise_colors_difference}
\end{figure}

Outflows that are detected through optical emission lines are expected to have large columns of gas, and/or large covering factors (otherwise they would not be detected in optical emission lines). These outflows must also contain dust (see discussion in \citealt{baron17b} and \citealt{baron18}), which is exposed to the AGN radiation and hence emits in the IR. Its temperature and peak wavelength depend on the bolometric luminosity of the AGN and the location of the wind with respect to the central source. Therefore, we expect to find differences in the IR SED of windy systems compared to the non-windy galaxies. We show in figure \ref{f:wise_colors_difference} a comparison of the WISE colors for the windy and the non-windy systems. One can see that windy systems are systematically redder than non-windy systems, except perhaps the case of W3-W2.

We examine the differences between the windy and non-windy systems in two different ways. The first is by using stacked SEDs. The advantage of this method is that intrinsic differences between sources, such as the torus inclination or the direct starlight properties, are averaged out. If all other properties are controlled for, any observed difference between the windy and non-windy SEDs can be attributed to the presence of an outflow (section \ref{s:stacked_analysis}). We also perform SED-fitting to individual systems (see section \ref{s:individual_objects}), and discuss the results of both methods in section \ref{s:sed_fitting_results}. 

\subsection{Stacked SEDs}\label{s:stacked_analysis}

The observed SED consists of contributions from a direct starlight component, torus and NLR dust emission, and emission of dust in SF regions. To control for these effects, we divide the objects in our sample into bins of stellar mass and SFR, thus ensuring that windy and non-windy systems will have the same direct starlight in the optical-NIR and the same SFR properties. For a given bin of stellar mass and SFR, we match the dust-corrected narrow [OIII] luminosity such that the windy and non-windy systems will show similar [OIII] distributions. The latter is done in order to make sure that the torus and NLR luminosities of the two groups are similar. This is based on the observation that both the torus and [OIII] average luminosities are directly related to the AGN bolometric luminosity (e.g., \citealt{netzer09}, \citealt{lusso11}, and discussion in \citealt{netzer13}). The matching is done by keeping all the objects that are located in the mutual region of the distributions. Outside the mutual region, the two distributions are not alike. To match the two distributions outside the mutual region, we sample objects from the larger distribution (i.e., the distribution with the larger number of objects) according to the distribution of the smaller group.

Table \ref{t:bins} in appendix \ref{a:stacks_bins} lists the bins and the number of objects in each bin. We also show in figures \ref{f:oiii_histogram_of_bins}, \ref{f:mass_histogram_of_bins}, and \ref{f:sfr_histogram_of_bins} the distributions of the narrow [OIII] luminosity, the stellar mass, and the SFR respectively, for the windy and non-windy groups. The diagrams show that within each bin the distributions of the windy and non-windy systems are similar in all these properties. Therefore, if the IR SEDs consist only of direct starlight, torus and NLR, and SF contributions, we do not expect to find any differences between the windy and non-windy SEDs. 

\begin{figure*}
\includegraphics[width=0.9\textwidth]{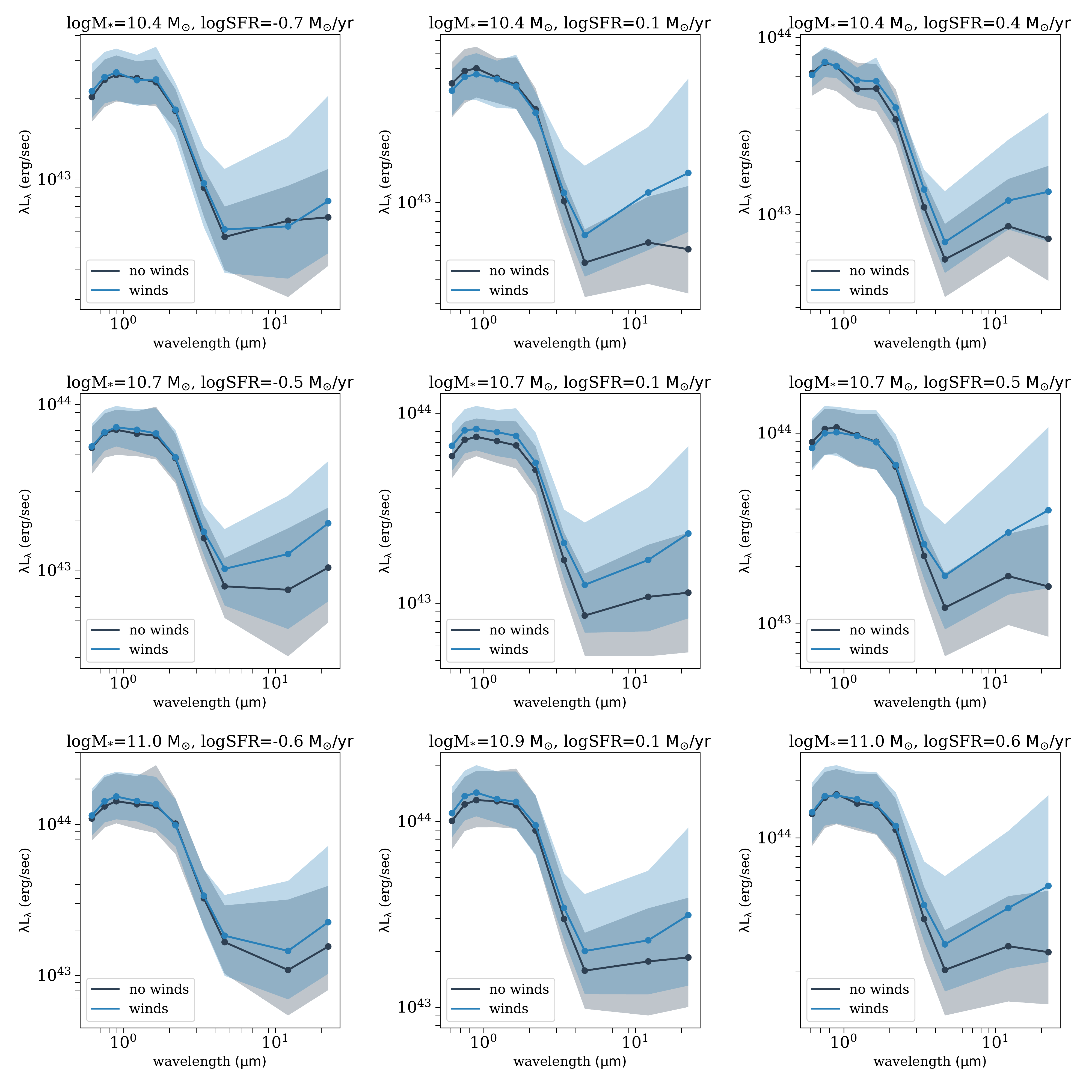}
\caption{\textbf{Comparison of stacked SEDs for the windy (blue) and non-windy (black) type II AGN.} We divide the objects into three bins of stellar mass and three bins of SFR. In each bin we select objects so that their narrow L([OIII]) distributions are similar. These ensure that the direct starlight component, the torus, and the SFR properties of the two groups are matched. Different rows correspond to different stellar mass bins, and different columns correspond to different SFR bins. The lines and points mark the median SED in each bin, while the shaded regions mark the 16th and 84th percentiles.}\label{f:stacked_SEDs_matched_oiii}
\end{figure*}

We compute the median and the 16th and 84th percentiles of the luminosities, as a function of wavelength, for each bin. We do not shift the wavelengths to restframe wavelengths since it requires making assumptions about the underlying SED. The low redshift of the sample ensures that the observed wavelengths are close to the restframe wavelengths. Furthermore, the central wavelengths of the broad bands are far enough from each other so that different bands do not overlap after redshift correction. Finally, the redshift distributions of the windy and non-windy samples are similar, and should not affect the stacks. The 16th and the 84th percentiles are computed directly from the luminosities, without taking into account the photometric uncertainties, since the intrinsic scatter is much larger than the individual uncertainties. We show in figure \ref{f:stacked_SEDs_matched_oiii} the comparison between the stacked SEDs of windy and non-windy galaxies for the different stellar mass and SFR bins. The blue and black lines represent the median SEDs of the windy and non-windy systems respectively, and the shaded regions mark the 16th to 84th percentiles. 

As evident from figure \ref{f:stacked_SEDs_matched_oiii}, in most of the bins, the windy systems show an excess emission with respect to the non-windy systems in the W3 and W4 WISE bands. Since the narrow L([OIII]) distributions of the two samples are similar, this difference cannot be due to the torus or the stationary NLR. Furthermore, the similar SFRs in the two cases suggests that this difference is not due to SF. We suggest that this excess is related to the presence of a dusty wind, where the dust is heated by the AGN, and causes excess emission at MIR wavelengths. Such an excess emission is unavoidable and the only remaining questions related to its detection at MIR wavelengths are the dust temperature and total luminosity. Under such circumstances, the NIR-MIR SED of an AGN with optically-detected outflows offers novel constrains on the wind properties, as we discuss below. The lowest stellar mass and lowest SFR bin in figure \ref{f:stacked_SEDs_matched_oiii} is the only bin in which we cannot distinguish between windy and non-windy systems. This can be related to the temperature of the dust in the wind. If the dust is colder than roughly 50 K, its emission will not be detected through the WISE bands. Alternatively, it can be related to a low covering factor of this dusty component, compared to the intrinsic scatter of torus properties in this bin.

\subsection{Individual SEDs}\label{s:individual_objects}

The stacked SEDs provide information about the general properties and differences between windy and non-windy systems. In this section we examine whether individual SEDs can be used to provide information about the properties of the outflowing dust component. 

We model each SED as a combination of a direct starlight in optical-NIR wavelengths, torus emission which already contains contribution from a stationary NLR, dust that is heated by SF, and an additional greybody component that represents the dusty wind. For the direct starlight component, we use the SWIRE\footnote{\url{http://www.iasf-milano.inaf.it/~polletta/templates/swire_templates.html}} template library (e.g., \citealt{polletta07}), which includes templates of 3 ellipticals, 7 spirals, 6 starbursts, 7 AGNs (3 type I AGNs, 4 type II AGNs), and 2 composite (starburst+AGN) templates covering the wavelength range between 1000\AA\, and 1000$\mathrm{\mu m}$. The elliptical, spiral and starburst templates were generated with the GRASIL code \citep{silva98}. Since we are not interested in the stellar population properties, and use only $riz$ and $JHK$ photometry, we modeled the SEDs only with the 3 elliptical templates, with ages of 2, 5, and 13 Gyrs. For each of the objects in our sample, we normalize the template to have a similar stellar mass to that reported by the MPA-JHU catalog, and select the template that best fits the photometry. Thus, the normalization of the template (which is connected to the stellar mass) is not a free parameter of the fit. 

We use the \citet{chary01} templates to model the SF contribution to the MIR\footnote{\url{http://david.elbaz3.free.fr/astro_codes/chary_elbaz.html}}. The library contains 105 template SEDs that were built to reproduce the observed luminosity-luminosity correlations for local galaxies. The SEDs cover the wavelength range 0.1 to 300\,000 $\mathrm{\mu m}$, and the library contains the integrated 8--1000 $\mathrm{\mu m}$ IR luminosity for each template. We use the Dn4000-based SFR provided by the MPA-JHU catalog, and convert it to 8--1000 $\mathrm{\mu m}$ luminosity (LIR) assuming $\mathrm{1\,M_{\odot}/yr = 10^{10}\,L_{\odot}}$ (see e.g., \citealt{netzer16} and \citealt{lani17}). We use this LIR to select the SF template for each object. The \citet{chary01} templates include also a contribution from direct starlight in the optical and NIR, which can differ, substantially, from the observed spectra. Since we account separately for the direct starlight component, we subtract from the SF template the contribution of the direct starlight using the SWIRE library. Thus, the final SF template includes only the FIR dust and MIR PAH emission. We subtract this template from the observed SED. Since the template is completely determined by the MPA-JHU SFR measurement, and is subtracted from the each observed SED, the SF template is not a free parameter of our fit.

For the torus emission, we use the dusty torus templates by \citet{stalevski12}\footnote{\url{https://sites.google.com/site/skirtorus/}}, with updates from \citet{stalevski16}. This library contains emission models of the AGN clumpy torus, calculated with the SKIRT radiative transfer code which is based on a Monte Carlo technique. The templates cover the wavelength range $10^{-3}$ to $10^{3}\, \mu m$. We use templates with average edge-on optical depth at 9.7 $\mu m$ of 5 and 10, a radial gradient of dust density of 1, and dust density gradient with polar angle of 2 (see \citealt{stalevski12} for details). Since our sample consists of type II AGN, we only use the templates in which the inclination of the torus with respect to the line of sight is in the range of 40--90 degrees. While the Stalevski et al templates describe only the torus emission, and not the NLR, they are successfully used to model the total MIR emission in type I AGN, which also have a contribution from the NLR (e.g., \citealt{duras17, ohyama17}). They are also similar to empirical MIR SEDs of type I AGN, which are designed to take the NLR dust into account (e.g., \citealt{mor12, netzer16, lani17}). Thus, we do not include an additional NLR component in our fit. In what follows, we assume that the NLR contribution over the wavelength range of 3--30 $\mathrm{\mu m}$ is $\sim$8\%, which is based on the empirical MIR template by \citet{mor12}.

The normalization of the torus luminosity requires additional assumptions since in type-II AGN the torus is not seen face-on and the self-absorption of the torus $\sim$1--5 $\mathrm{\mu m}$ radiation prevents us from directly measuring the part of the AGN optical-UV radiation absorbed and then re-emitted at IR wavelengths. To estimate this, we apply the following procedure. We use the dust-corrected narrow [OIII] and H$\beta$ luminosities to estimate the bolometric luminosity of the AGN (\citealt{netzer09}): 
\begin{equation}\label{eq:1}
	{\log L(\mathrm{AGN}) = \log L(\mathrm{H}\beta) + 3.48 + \mathrm{max \Big[ 0, 0.31\big( \log \frac{[OIII]}{H\beta} -0.6 \big) \Big] }}
\end{equation}
Since the dusty torus absorbs the radiation originating in the accretion disk, the bolometric luminosity of the AGN and the total IR luminosity of the torus are connected via the torus covering factor. For typical low redshift, type I and type II AGN, the integrated torus luminosity is estimated as $\mathrm{\nu L_{\nu, torus}(20\mu) \sim 0.13 L(AGN)}$ (e.g., \citealt{netzer13}). Since this is an empirical estimate based on the total torus and (much smaller) NLR dust emission, it provides a way to estimate the torus normalization using the reddening-corrected [OIII] and H$\beta$ luminosities. The relation between the observed emission line luminosities ([OIII] and H$\beta$) and the torus emission, through the bolometric luminosity, shows a non-negligible scatter due to the different covering factors of the torus, the covering factor of the NLR (L([OIII]) and L(H$\beta$)), and the AGN luminosity (see e.g. \citealt{netzer09, mor09, lusso11, mor12, ichikawa17}). We therefore allow the normalization of the template to be within a factor of 1/6 to 6 from our initial estimated value. Therefore, the two free parameters in our fit are the inclination-dependent torus template and the torus normalization, where the latter is limited to the noted range. 

Finally, we model the additional dusty wind component as a single greybody, with temperatures ranging from 30 K to 1\,000 K and $\beta = 1.5$. The lower limit is chosen to roughly correspond to the coldest dust component we can measure, and is set by our longest accessible wavelength of 22 $\mathrm{\mu m}$. The upper limit is chosen to correspond to the hottest dust temperature we can detect, given that the direct starlight component dominates the SED at NIR wavelengths. The normalization of the greybody is a free parameter of our model. 

All the templates described above are summed to obtain the combined SED for a given system. Contrary to the stacked analysis, here we shift the templates according to the redshift of each object. We then perform synthetic photometry of the predicted SED and minimize the residuals between the synthetic photometry and the observed photometry. We perform two SED fits for each object in our sample: with and without the additional dusty wind component. For the first fit we have 3 free parameters, one for the direct starlight component and two for the torus. For the second fit we have additional 2 free parameters for the greybody component. Since we have 10 photometric measurements for each object, we have 7 and 5 degrees of freedom respectively. The photometric uncertainties reported for the SDSS, 2MASS, and WISE bands are usually less than 1\% of the measured value. Given the uncertainty in the various SED templates and the other assumptions of the model, and systematic uncertainties in the various inter-band calibration, we consider this unrealistic. We therefore added additional 10\% of the measured value to the error. 

To evaluate the goodness of fit we define a standard score $R$ which is applied to the two cases, with and without additional greybody component:
\begin{equation}\label{eq:0}
	{R = \frac{1}{M} \sum\limits_{i=1}^n \frac{(y_{i} - f_{i})^2}{\sigma_i^2}}
\end{equation}
where $M$ is the number of degrees of freedom, which equals 7 for the case without the greybody and 5 for the case with the greybody component. The measured luminosity is $y_{i}$, and the synthetic luminosity of the best-fitting template is $f_{i}$. As explained, the uncertainties $\sigma_{i}$ take into account the very small flux uncertainties and the larger uncertainties on the template SED. Thus, $R$ is not equivalent to the standard $\chi_{red}^{2}$ statistic. Nevertheless, $R$ can be used as a goodness of fit measure. 

For a given object, we examine $R_{\mathrm{wind}} = R(\mathrm{no\,GB}) - R(\mathrm{with\,GB})$, where $R(\mathrm{no\,GB})$ is the score for the model that does not include the greybody, and $R(\mathrm{with\,GB})$ corresponds to the model that includes the greybody. We expect this value to be larger than 0 for objects in which an additional greybody component is required by the fit, and below 0 otherwise. Using the optical classification of windy versus non-windy objects, we also expect that systems that do not show  wind components in their optical spectra will not require an additional greybody component, while systems where outflows are detected will require an additional component.

\begin{figure}
\includegraphics[width=3.25in]{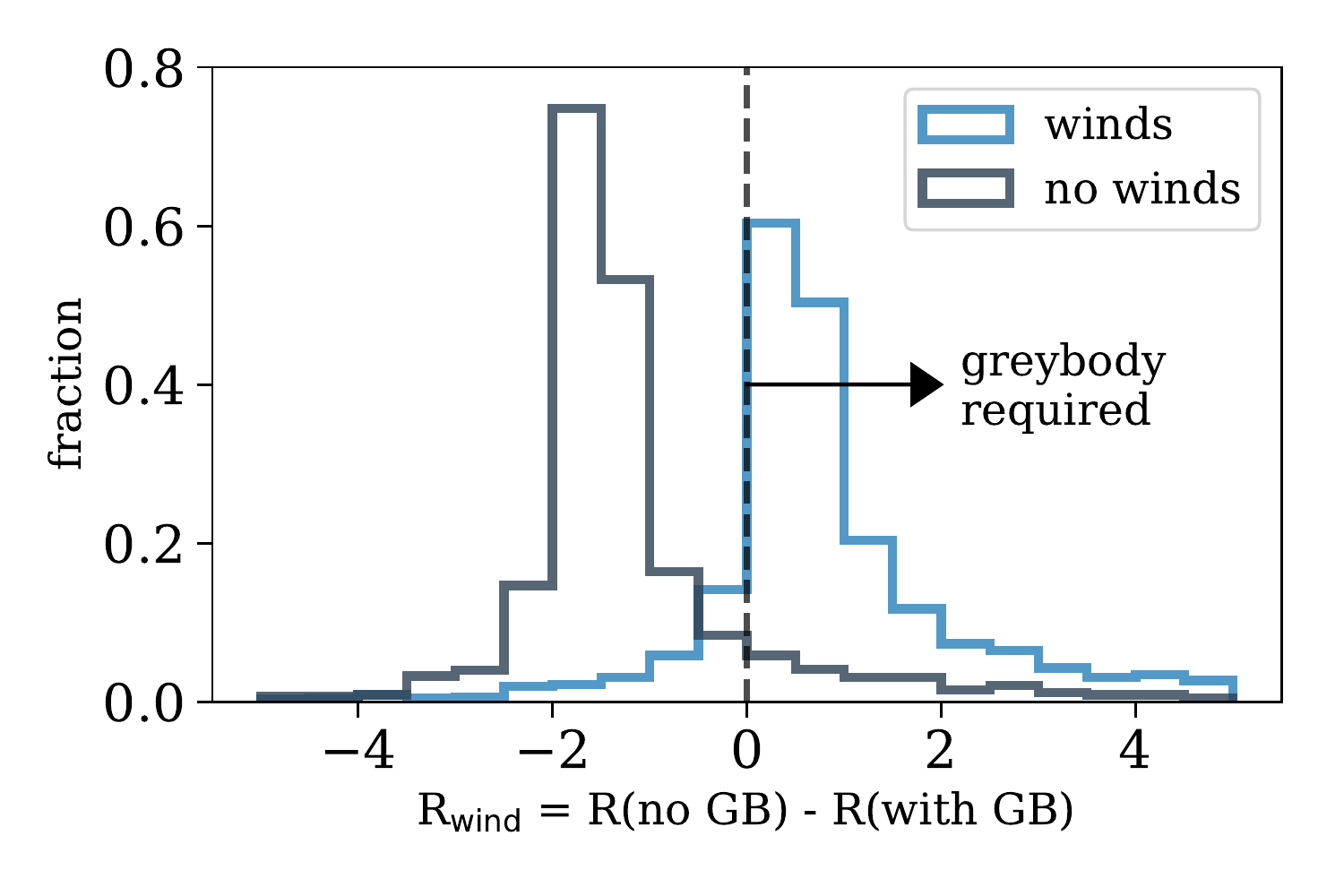}
\caption{A comparison of the distribution of $R_{\mathrm{wind}} = R(\mathrm{no\,GB}) - R(\mathrm{with\,GB})$ for the windy and non-windy objects. $R(\mathrm{no\,GB})$ is the score for models that do not include a greybody component, and $R(\mathrm{with\,GB})$ corresponds to models that include a greybody component. A threshold of $R_{\mathrm{wind}} = 0$ is used to determine whether the best model includes or does not include a greybody component.}\label{f:scores_comparsion_models}
\end{figure}

We show in figure \ref{f:scores_comparsion_models} the distribution of $R_{\mathrm{wind}}$ for the windy and non-windy systems. The figure shows that the windy and non-windy systems are well-separated when performing individual SED fitting, compared to the stacked SEDs (figure \ref{f:stacked_SEDs_matched_oiii}) where a considerable overlap is observed. We attribute this difference to the large range in [OIII] luminosities in each of the bins used for the stacks (1-2 orders of magnitude; see figure A1 in the appendix). That is, the overlap in the stacked SEDs is due to the intrinsic variations in torus properties, which is better controlled for in the individual SED fits. As expected, systems that show outflows through their optical emission lines also show $R_{\mathrm{wind}} > 0$ in most cases, and systems that do not show spectroscopic outflow signatures show usually the opposite. Since we expect that no additional greybody component will be required for the non-windy systems, we choose a threshold of $R_{\mathrm{wind}} = 0$ to determine which model is chosen for a given object. We find that 81\% of the non-windy systems do not require an additional greybody component. We also find that 86\% of the windy systems require an additional greybody component. However, as we discuss in section \ref{s:wind_properties}, we cannot constrain the greybody luminosity in 347 of the systems. Thus, we are able to constrain the dust properties only for 71\% of the windy systems.

\begin{figure*}
\includegraphics[width=0.85\textwidth]{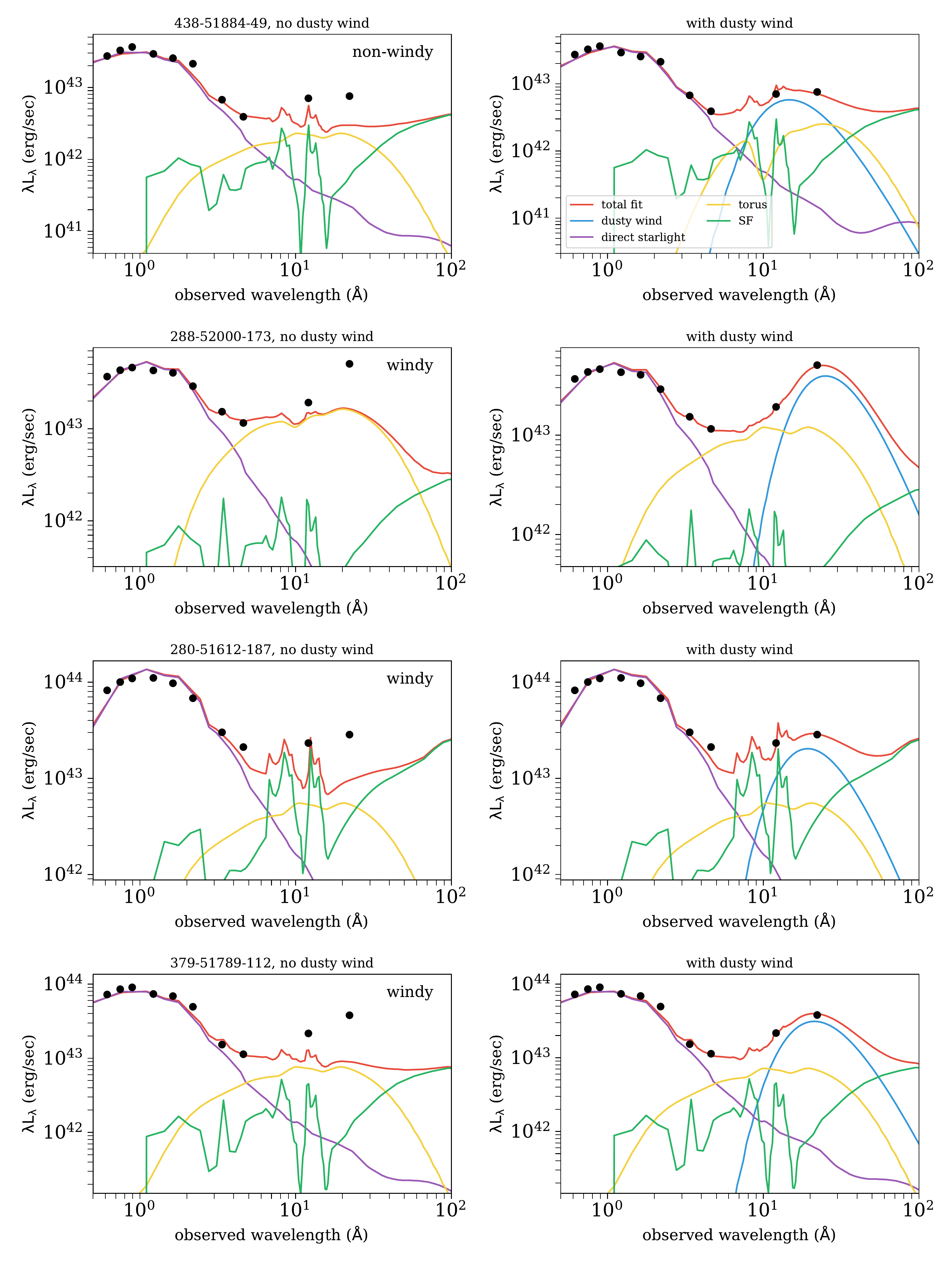}
\caption{Four examples of best-fitting SEDs for cases in which an additional dusty wind component is required. Measurements are marked with black points, the best-fitting SED with red, the direct starlight component with purple, the SF template with green, the torus with yellow, and the additional dusty component with blue. Each row represents a galaxy (with its SDSS plate-mjd-fiber noted in the title). The left column shows the best fit where we do not include the additional dusty component while the right column shows the best fit which this dust. The insert labels in the left column indicate whether the galaxy is classified as windy or non-windy from its optical emission lines.}\label{f:fit_examples_with_gb}
\end{figure*}

\begin{figure*}
\includegraphics[width=0.85\textwidth]{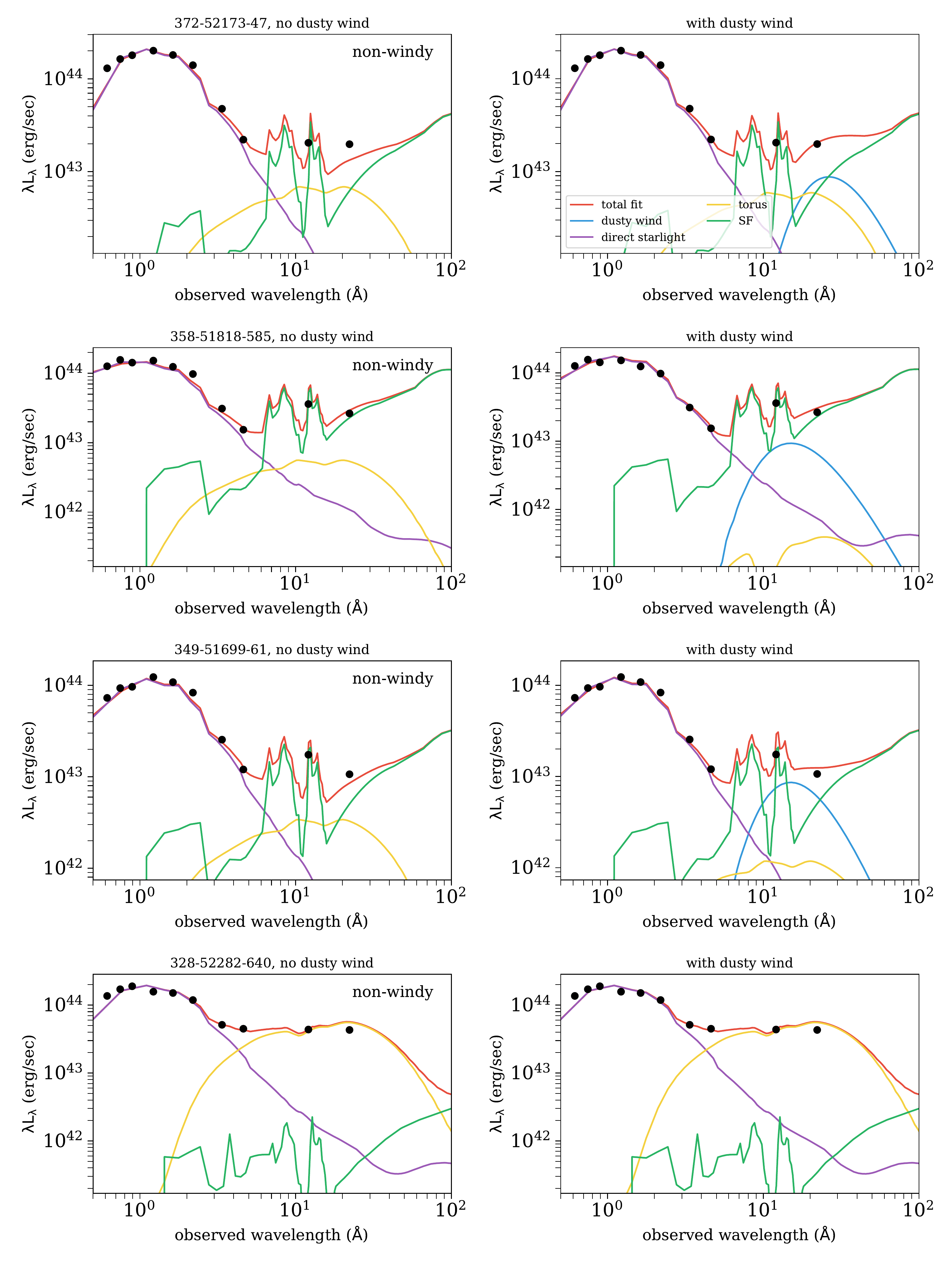}
\caption{Same as figure \ref{f:fit_examples_no_gb} but for cases in which an additional dusty wind component is not required. The insert labels in the left column indicate whether the galaxy is classified as windy or non-windy from optical emission lines.}\label{f:fit_examples_no_gb}
\end{figure*}

We show in figure \ref{f:fit_examples_with_gb} four examples of objects for which the chosen model includes a greybody component. The measured luminosities are marked with black points, the best-fitting SED in red, and separate templates are shown with different colors. Each row represents a single object, where the left panel shows the best-fitting SED in the case where the dusty wind is not included, and the right panel shows the best fit when the additional dusty component is included. The first object in this group, shown in the top row, does not show a spectroscopic signature of a wind, yet a wind MIR component clearly improves the fit. This object belongs to the 19\% of the non-windy systems for which $R_{\mathrm{wind}} > 0$ (see figure \ref{f:scores_comparsion_models}). The three additional sources clearly show a spectroscopic signature of a wind and the SED fits that contain an additional greybody are superior to those where such a component is not included. In figure we show \ref{f:fit_examples_no_gb} four examples of objects with no spectroscopic signature of a wind, and for which the chosen model does not include a greybody component. As expected for such systems, the fit without the dusty wind is good enough and such a component is not required.

\subsection{Torus and NLR properties}\label{s:sed_fitting_results}

The results presented above show that the two methods used here, stacked SEDs and individual SED-fitting, both indicate that systems with spectroscopic indications for an ionized gas outflow are better fitted with a model that includes an additional greybody component. The other properties of the two groups, SFR, stellar mass, and dust-corrected [OIII] luminosity, are very similar and do not depend on the additional wind component. This suggests that the difference is not driven by the SF or torus properties of the two populations. 

Next, we examined the possibility that a difference in the gas properties in the NLR drives the observed MIR differences. Since the dusty NLR gas is exposed to the radiation originating in the accretion disk, different gas properties in the two groups will result in different dust properties, which may account for the observed MIR excess. To check this, we first compare the reddening, $\mathrm{E}(B-V)$, towards the narrow lines in the two groups. We find no significant difference in dust reddening between the windy and non-windy systems. 

We also examined the narrow [OIII]/H$\beta$ line ratio, which is related to the degree of ionization of the NLR gas. We find that windy systems have a slightly larger [OIII]/H$\beta$ ratio than the non-windy systems, with the means of the distribution log([OIII]/H$\beta$)$\sim 0.8$ for the windy systems and log([OIII]/H$\beta$)$\sim 0.6$ for systems without winds. This suggests that the NLR gas in windy systems is slightly more ionized. For a given ionizing source luminosity, higher ionization suggests that the gas is either closer to the central source, or its density is lower, compared to a lower ionization gas. If the first explanation is correct, then the dust temperature in the NLR of windy systems is higher.

There are two reasons why the differences in [OIII]/H$\beta$ between windy and non-windy systems cannot explain the excess MIR emission. First, as already alluded to, all AGN have dusty NLRs that contribute to the observed MIR luminosity \citep{schweitzer08, mor09, mor12}. If the observed MIR excess in windy systems is due to more NLR dust, there is also more narrow line reddening in such systems. This contradicts our finding of no difference in reddening between windy and non-windy systems. Second, as we show in section \ref{s:wind_properties} below, the MIR excess is about 10\% of the AGN bolometric luminosity. Such a large excess cannot be explained by a somewhat higher NLR dust temperature. On the other hand, the broad [OIII] luminosity, which is related to the optically-detected outflow, with luminosity which equals or even exceeds L([OIII] narrow), is entirely consistent with the observed MIR excess.

Next, we compare windy systems with detected greybody component (86\% of the objects) to windy systems where such a component was not detected (14\% of the objects). We compared the SFR, stellar mass, and dust corrected narrow [OIII] luminosity, and found no differences between the two subgroups. A non detection of the greybody component can be due to its relative weakness compared to the torus emission, since the two radiate most of their emission at roughly the same wavelength ranges. The narrow [OIII] luminosity represents the bolometric luminosity and thus the torus luminosity, assuming typical covering factors for both components. The broad [OIII] luminosity represents the total luminosity of the greybody component. Thus, we expect to find a difference in the ratio of the two line components if the non detection is due to a relative weakness of the greybody with respect to the torus. We find no such difference between the two groups. We carried out similar tests for the non-windy systems and found, again, no such differences.

We use the best-fitting SEDs to estimate the torus covering factor in the windy and the non-windy systems. As mentioned previously, integrating over the torus template in type II AGN will not recover the total disk energy absorbed by the torus, due to self-absorption in the torus. Thus, we apply the following procedure. We use the best SED fit to obtain $\mathrm{\nu L_{\nu} (20 \mu m)}$ for the windy and non-windy systems. We then integrate the type I AGN torus template given by \citet{mor12}, and obtain the relation between the total IR luminosity of the torus and NLR and $\mathrm{\nu L_{\nu} (20 \mu m)}$, which is $\mathrm{L(torus)} = 3.72 \times \mathrm{\nu L_{\nu} (20 \mu m)}$ (the \citealt{stalevski12} type I template gives a factor of 3.57). The emitted $\mathrm{20\, \mu m}$ radiation originates in the outer region in the torus, by a colder dust, which is not affected by scattering and absorption within the torus. This luminosity is assumed to be similar in type I and type II AGN.

\begin{figure}
\includegraphics[width=3.25in]{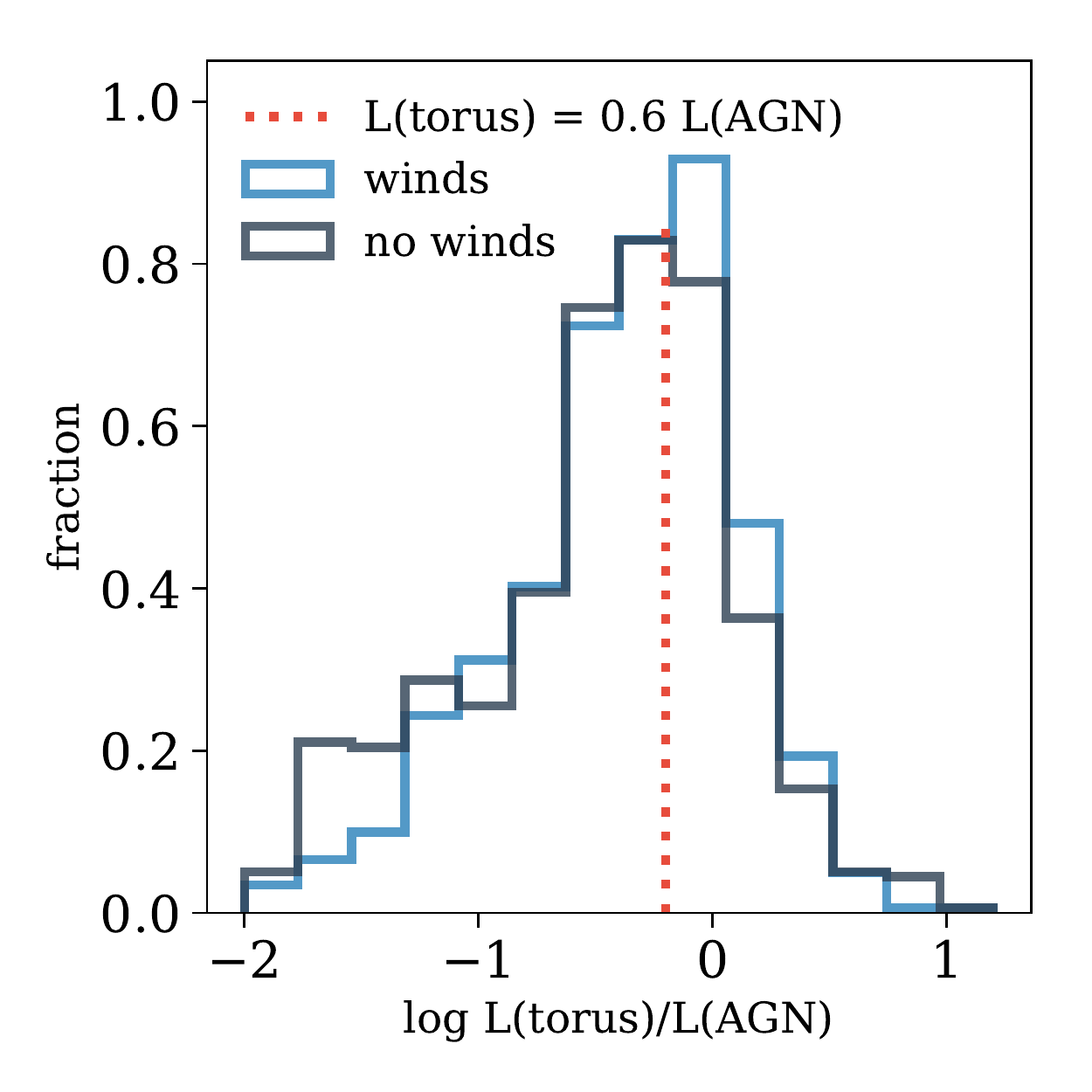}
\caption{The distribution of $\mathrm{L(torus)/L(AGN)}$ for the windy and non-windy objects. We estimated L(AGN) using the dust-corrected narrow emission lines, and L(torus) using the torus normalization from the best-fitting SED. The two distributions are very similar, and peak around $\mathrm{L(torus)/L(AGN)=0.6}$, which is consistent with estimates of torus covering factors in type I AGN.}\label{f:torus_covering_fraction}
\end{figure}

We show in figure \ref{f:torus_covering_fraction} the distribution of the torus covering factor for windy and non-windy systems. The two distributions are similar and both peak at $\mathrm{L(torus)/L(AGN) \sim 0.6}$. This provides a consistency check of our SED fitting procedure. Following \citet{netzer16} we deduce torus covering factor of 0.6 (0.38) for the isotropic (anisotropic) emission case. This estimate of the covering factor is consistent with other estimates in type I AGN \citep{mor09, mullaney11, mor12, lusso13, netzer16, stalevski16, lani17}. We note, again, that while prior to the SED fitting we estimated the torus luminosity from the narrow [OIII] and H$\beta$ emission lines, the range of normalizations was very large (from six times smaller to six times larger of the estimated value). Therefore, finding a consistent covering factor suggests that our SED fitting accounts correctly for the torus emission. One can see, however, that the scatter around the mean is large, which is to be expected given the uncertainties involved in the SED fitting and in estimating L(AGN) from the narrow line luminosity.

\begin{figure*}
\includegraphics[width=0.95\textwidth]{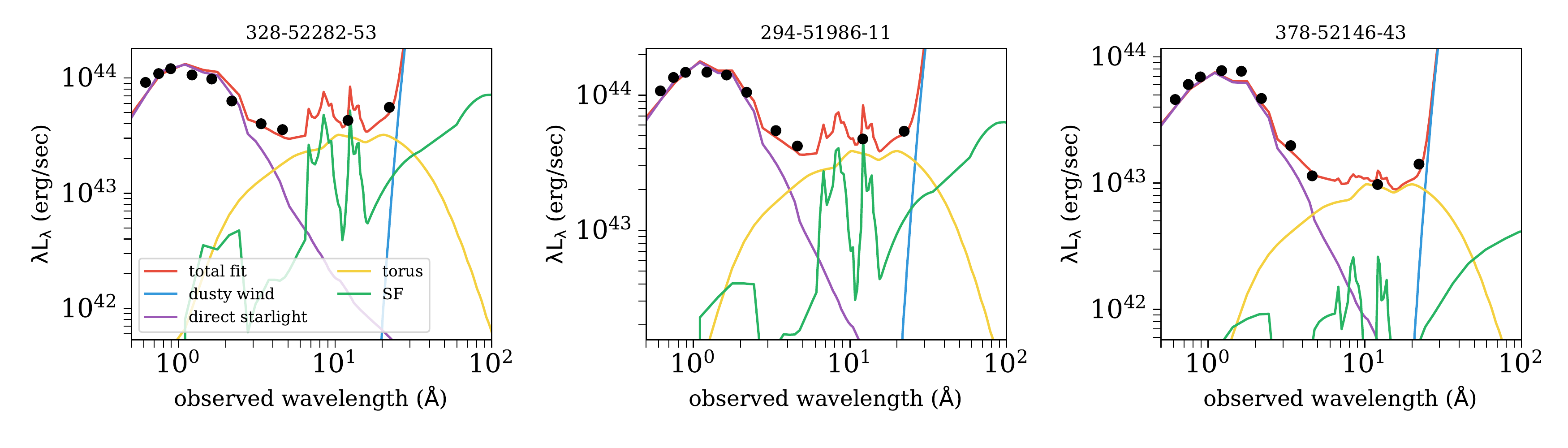}
\caption{Three examples of best-fitting SEDs that result in a greybody luminosity which exceeds the torus luminosity by more than an order of magnitude. The greybody component is required to account for an excess in the W4 band, but not the other bands. Clearly, the normalization and the temperature of the greybody component are unconstrained in such cases.
}\label{f:fit_examples_with_gb_bad_objects}
\end{figure*}

\section{Wind properties}\label{s:wind_properties}

Out of the 2\,377 systems with spectroscopic detection of a wind, 2\,043 systems require a greybody wind component. Out of these, there are 347 systems in which the greybody luminosity exceeds the torus luminosity by more than an order of magnitude, with a greybody temperature of 30 K. We show three examples of such systems in figure \ref{f:fit_examples_with_gb_bad_objects}. In these systems, a greybody component is required in order to account for an excess in the W4 band, but not in the other bands, and thus it improves the overall fit. However, as can be seen clearly in figure \ref{f:fit_examples_with_gb_bad_objects}, the normalization of the greybody is unconstrained, and the dust temperature is uncertain. In what follows, we remove these objects from the sample (the stacks in figure \ref{f:stacked_SEDs_matched_oiii} do not change when we remove the 347 objects). We examine the temperatures of the outflowing dust in section \ref{s:dust_temperature}, estimate the mean location of the wind in section \ref{s:wind_location}, and discuss the wind covering factor in section \ref{s:wind_covering_factor}.

\subsection{Dust temperature}\label{s:dust_temperature}

\begin{figure}
\includegraphics[width=3.25in]{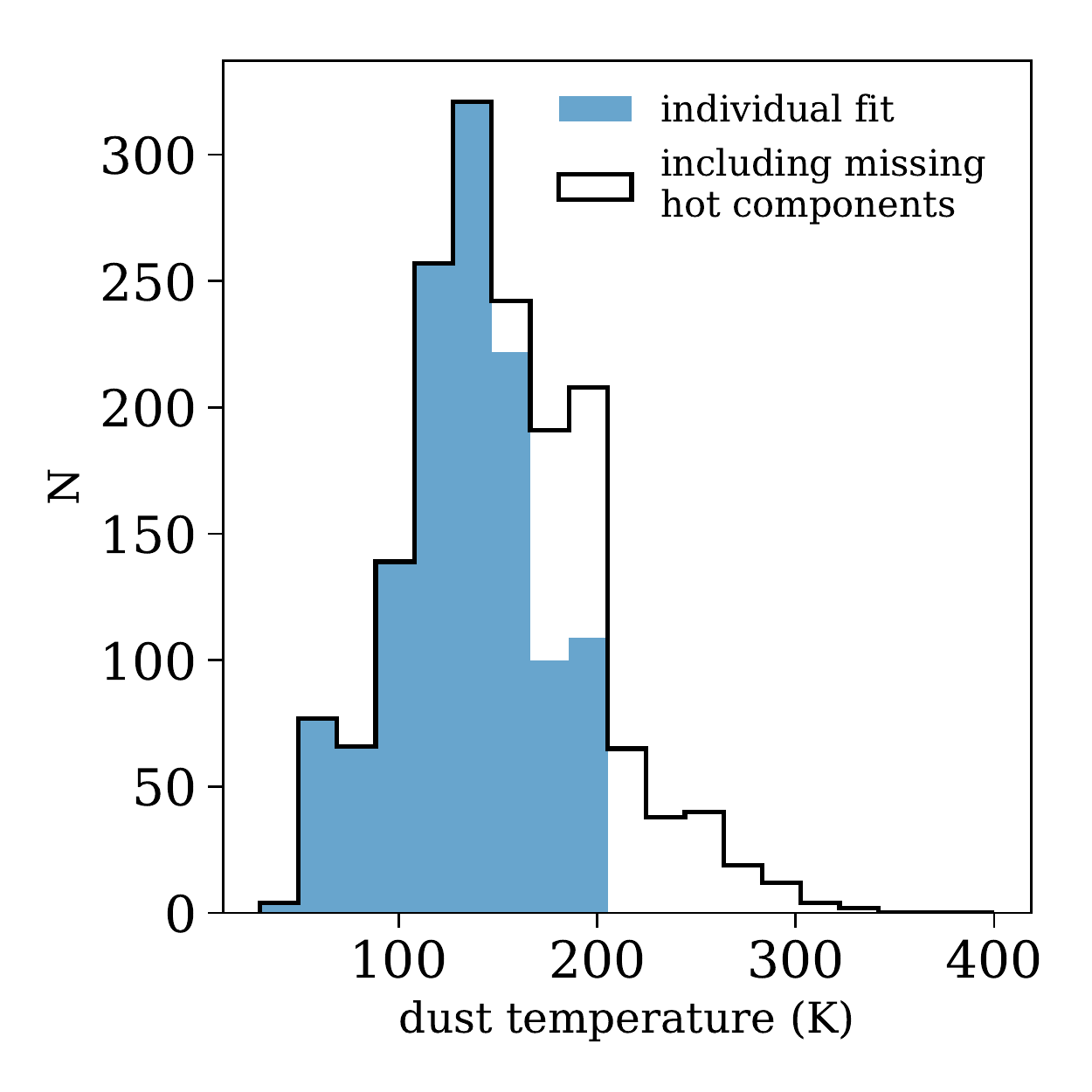}
\caption{The distribution of dust temperature obtained by SED-fitting to individual windy objects (blue). This distribution is compared to the dust temperature distribution that is required to fully account for the observed MIR excess in the stacked spectra (white). The latter contains a contribution from a hotter dust component, which is not detected in individual systems due to degeneracy with the torus dust. The hotter dust component may be related to the windy systems in which we do not detect an additional greybody component (14\%), or can be related to systems that host several phases of dusty wind, in which we detected only the colder component. The total distribution represents the general dust temperature distribution in AGN outflows.
}\label{f:dust_temperature_distributions}
\end{figure}

The outflow component in the IR SED fit provides an estimate of the dust temperature. We show in figure \ref{f:dust_temperature_distributions} the distribution of the dust temperature obtained from individual SED fits to windy systems in our sample. The distribution does not include the 347 objects removed in the previous stage. The distribution should be consistent with the net MIR excess we observe in the stacks. To test this, we calculated the median SED of this dusty component. The luminosity of this dusty component is roughly constant with respect to the luminosity of the torus. We can therefore obtain the median synthetic SED by summing grey bodies with the temperature distribution shown in figure \ref{f:dust_temperature_distributions}, where all greybody components have the same luminosity. By comparing this median dust SED to the observed excess in the stacks (figure \ref{f:stacked_SEDs_matched_oiii}), we find a good agreement at longer wavelengths (W3 and W4). However, the median synthetic emission under-predicts the observed excess at shorter wavelengths (W1 and W2; see e.g. bottom right panel of figure \ref{f:stacked_SEDs_matched_oiii}), which correspond to higher dust temperatures. This suggests that our SED-fitting procedure is not sensitive enough to detect hotter dust, which is not surprising since the hot dust emission covers a wavelength range similar to the one covered by the torus.

Alternatively, we can check what type of dust temperature distribution will produce the observed luminosity excess between the windy and non-windy stacks. We add a population of sources with hotter dust outflowing components that, together with the dust temperatures we have detected, will completely account for the MIR excess observed in the stacks. We show in figure \ref{f:dust_temperature_distributions} the dust temperature distribution that fully accounts for the observed excess. One can see that while our fitting procedure successfully detects dust temperatures below 200 K, it is not sensitive enough to detect higher dust temperatures. This behavior is consistent with the degeneracy between the torus and the additional hotter dusty wind component, and cannot be improved given the limited wavelength resolution that we use.

The population of sources with hotter outflowing dust cannot be directly compared to the windy systems for which a greybody component is not required (14\% of the systems). The presence of a population of hotter outflowing gas is inferred from the stacked SEDs, and cannot be traced to individual objects. These sources might be part of the 14\% undetected sample, but they can also be part of the 86\% sample for which we detect the greybody component. In the latter case, galaxies can host two outflowing phases, one which corresponds to a colder dust component which our method detects, and one corresponding to a hotter dust component, which we do not detect due to the degeneracy with the torus. Thus, the total temperature distribution in figure \ref{f:dust_temperature_distributions} (blue + white) represents the general distribution of dust temperatures in outflows. 

Finally, we emphasize that both the individual SED fitting procedure and the analysis of the stacked SEDs are not sensitive to dust temperatures below about 30 K, due to the limited wavelength range that we use (W4 is the longest wavelength band in our analysis). Additional FIR information is required to increase the sensitivity of the SED fitting method to detect colder dusty wind components. 

\subsection{Wind location}\label{s:wind_location}

\begin{figure}
\includegraphics[width=3.25in]{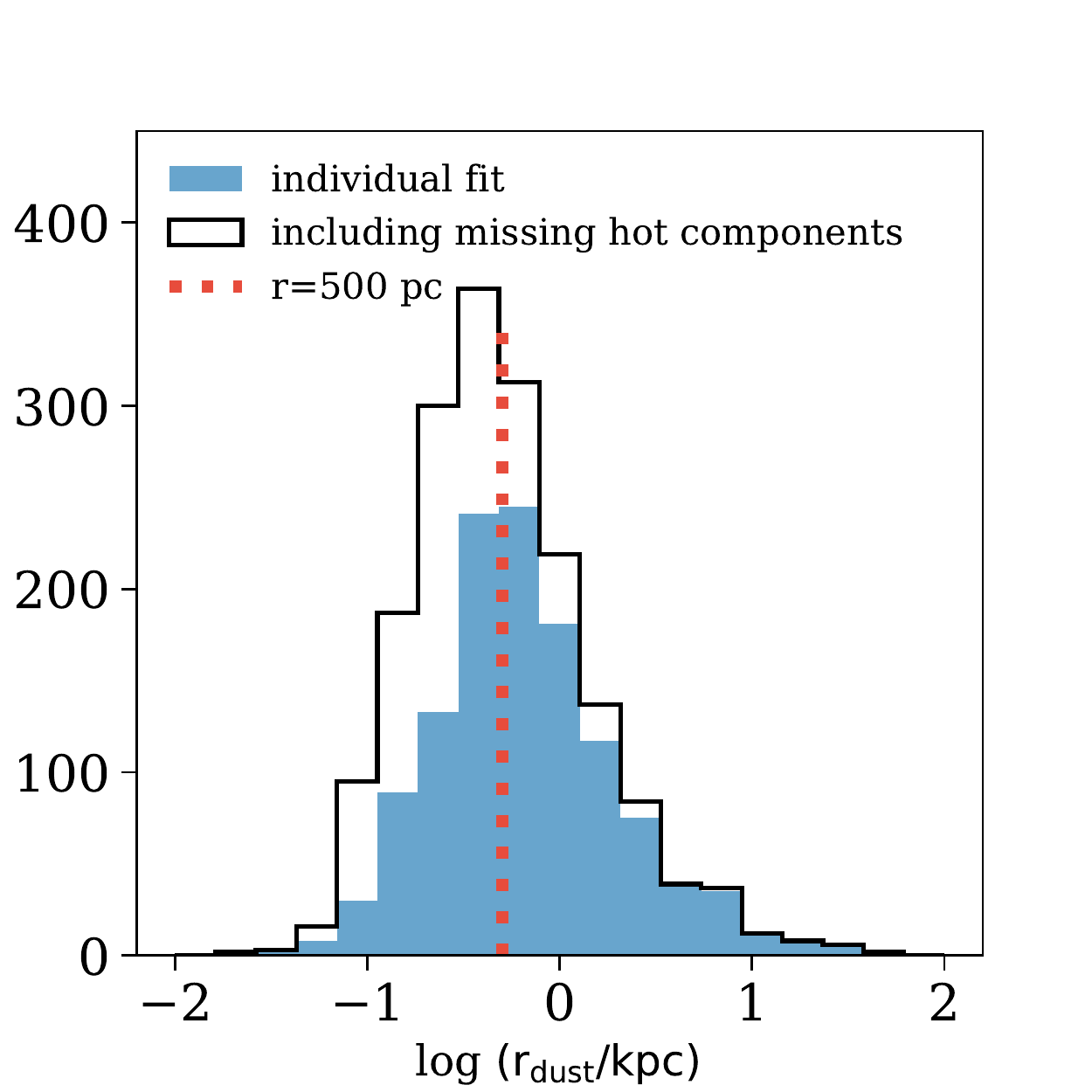}
\caption{The distribution of the location of the dusty component with respect to the central AGN. We use the best-fitting dust temperatures from the individual SED fitting procedure together with the bolometric luminosity of each system to estimate the dust location (blue). We estimate the location distribution of the entire windy sample (detected and not detected through IR SED fitting) using a median bolometric luminosity of $L(\mathrm{AGN}) \sim 10^{44.8}\,\mathrm{erg/s}$ and the temperature distribution from figure \ref{f:dust_temperature_distributions}. The total (blue + white) distribution represents the general distribution of outflow locations in active galaxies.}\label{f:dust_location_distributions}
\end{figure}

We use the dust temperatures from the SED fitting and the estimated bolometric luminosities to estimate the mean distance between the AGN and the dusty component. For dust which is optically thin for its own radiation, the distance from the central AGN is given by (e.g., \citealt{netzer13}):
\begin{equation}\label{eq:2}
	{\mathrm{\Big( \frac{r}{pc}\Big) = \Big( \frac{1500\, K}{T_{dust}}\Big)^{\frac{4 + \beta}{2}} \Big( \frac{L(AGN)}{10^{46} \, erg/s}\Big)^{\frac{1}{2}}}}
\end{equation}
where $\beta \sim 1.5$. We use equation \ref{eq:2} to calculate the distribution of distances for all type II AGN with dusty wind components discovered by our SED fitting process. We show the distribution in figure \ref{f:dust_location_distributions} (blue histogram). The peak of the distribution is at a distance of roughly 500 pc from the central source. We found in section \ref{s:dust_temperature} that our SED fitting procedure is likely to miss a hotter dust component, due to the degeneracy with the torus emission. We use the estimated temperature distribution in figure \ref{f:dust_temperature_distributions} which contains the hotter dust (the white histogram) to produce a distribution of dust locations for the entire sample. For this, we use a bolometric luminosity of $\mathrm{L(AGN)} \sim 10^{44.8}\,\mathrm{erg/s}$, which is the median bolometric luminosity in our sample. The estimated location distribution for the entire sample is shown in figure \ref{f:dust_location_distributions} (white histogram). One can see that the additional hotter dust component corresponds to smaller distances between the AGN and the dusty gas. 

Figure \ref{f:dust_location_distributions} shows a Gaussian-like distribution with a clear peak at around 500 pc. The distribution is somewhat asymmetric, with a longer tail towards larger wind extents, around $\sim$10 kpc. One can see that there are no objects below a dust location of roughly 50 pc. The distribution in figure \ref{f:dust_location_distributions} is somewhat inconsistent with wind locations reported by ground-based seeing-limited IFU studies, which report much larger wind extents (see e.g., \citealt{harrison14}). This may be partially related to differences in the definition of the wind location, which we further discuss in section \ref{s:mass_outflow_rate}. The distribution is more consistent with wind location estimates that are based on HST observations, or studies that correct for beam-smearing in seeing-limited observations (e.g., \citealt{husemann16,villar_martin16, fischer18, tadhunter18}).

We showed in section \ref{s:sed_fitting_results} that our SED fitting procedure cannot detect dust emission for dust temperatures lower than 30--40 K, since our reddest band is the W4 band. This means that for the median bolometric luminosity in our sample, we cannot detect winds through IR SED fitting to extents that are larger than about 5--10 kpc. Nevertheless, the winds were initially detected through optical emission lines using a 3'' fiber, which for a median redshift of $z=0.1$ covers 5.4 kpc of the host galaxy in diameter. These winds must be within 2.7 kpc from the center of the galaxy, otherwise they would not have been detected through the optical emission lines. Therefore, while it is clear that the SED fitting misses winds with large extents, the location distribution of the winds cannot be very different from what we show in figure \ref{f:dust_location_distributions}. We therefore suggest that the distribution of wind locations in figure \ref{f:dust_location_distributions} roughly represents the true distribution of type II AGN for which outflows are detected through optical emission lines for which $\mathrm{L(AGN)} \sim 10^{44.8}\, \mathrm{erg/sec}$. Obviously, there can exist a population of winds that have large extents, and even reside outside their host galaxy (see e.g. \citealt{baron18}), which will neither be detected in SDSS optical spectra nor through the IR SED fitting described here. We return to this point in section \ref{s:disc}.

We looked for other possible correlations with the wind location. We find no correlation between the location of the dusty wind and the stellar mass or the SFR in the galaxy. We also examined the relation of the location of the wind with the SF-to-AGN luminosity ratio, and find no correlation. Furthermore, we find no correlation between the location of the wind and its covering factor (see below). We do not examine the correlation of the wind location and the AGN bolometric luminosity, since the location is partially determined from L(AGN).

\subsection{Wind covering factor}\label{s:wind_covering_factor}

\begin{figure}
\includegraphics[width=3.25in]{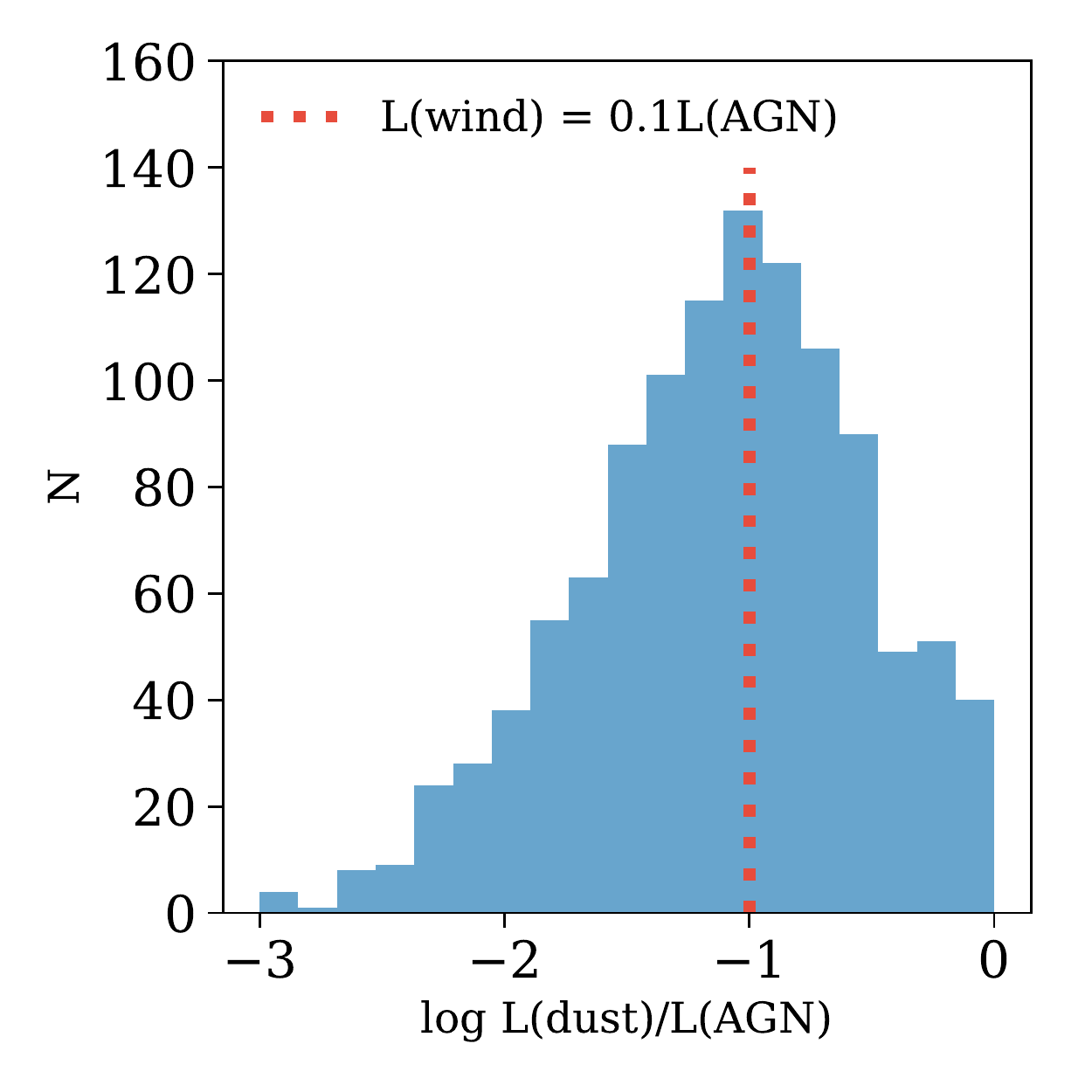}
\caption{L(dust)/L(AGN) for the sample of type II AGN with dusty winds. The distribution peaks at a covering factor of 10\%, with a larger scatter around it. We find a wind covering factor larger than 20\% (50\%) for 24\% (18\%) of the objects.}\label{f:dust_covering_factor}
\end{figure}

We use the properties of the dusty winds to estimate the covering factor of this component (note, again, the definition of covering factor explained in the introduction). We estimate the total IR luminosity of this component by integrating over the corresponding greybody, with the best-fitting normalization and dust temperature. We also assume that the optical depth of the dust in the outflow is large at optical-UV wavelengths and small at NIR-MIR wavelength; an assumption which is consistent with the column of ionized gas required to explain the intensity of the broad components in the profiles of the [OIII] lines. Under these assumptions, the wind covering factor is approximately the ratio of the dust luminosity to the total AGN luminosity. We show in figure \ref{f:dust_covering_factor} the distribution of L(dust)/L(AGN) for the windy systems in our sample, which represents the wind covering factor. The distribution peaks at a covering factor of 10\%, but shows a large scatter around this mean, with covering factors ranging from 0.1\% to 100\%. We find that 24\% of the winds cover more than 20\% of the central source, and that 8\% cover more than 50\%. Since this dust is mixed with the gas, the distribution roughly represents the covering factor of the ionized wind. We also note that the peak of this distribution is roughly twice the mean covering factor of the NLR (5\%; e.g. \citealt{mor12}). 

We find no correlation between the wind covering factor, CF(wind), and the stellar mass or SFR in the galaxy. We do find a correlation between CF(wind) and the SF-to-AGN luminosity ratio ($\rho = 0.55$). However, this correlation is due to the mutual dependance of CF(wind) and L(SF)/L(AGN) on L(AGN), and remains when we shuffle the L(AGN) values with which we measure CF(wind) and L(SF)/L(AGN). Thus, this correlation is not physical.

\section{Discussion}\label{s:disc}

By comparing the NIR-MIR SED of type II AGN with and without ionized outflows, we discovered an additional MIR emission component in the windy systems. We suggested that this emission is due to dust that is mixed with the gas in the outflow, and is heated by the AGN. We have shown that this component can be measured in many cases through SED fitting of individual objects, and explored the limitations of the method for different dust temperatures. We have also shown that, when detected, this dust component offers novel constraints on the wind properties, such as its mean location and covering factor. In this section we discuss how these results can be used to obtain better estimates for the mass outflow rates of winds in AGN (section \ref{s:mass_outflow_rate}). We then discuss the main uncertainties involved in the SED fitting and suggest ways to reduce them in section \ref{s:sed_fitting_uncertainties}.

\subsection{Mass outflow rate estimate from 1D spectra and photometry}\label{s:mass_outflow_rate}

The mass outflow rate and the kinetic power of the wind are fundamental properties in active galaxies, partly since they are directly related to AGN feedback \citep{silk10, zubovas14, croton16, wilkinson18}. The mass outflow rate can be estimated using a combination of several other observables: (1) dust-corrected luminosity of an emission line (typically H$\alpha$ or [OIII]) that is associated with the wind, (2) the wind velocity (emission line dispersion or shift, or a combination of the two; see \citealt{karouzos16b} for details), (3) the electron density in the outflowing gas, and (4) the location of the wind with respect to the center of the galaxy. The first two are rather straightforward to obtain from 1D spectra. The electron density can be estimated from the [SII] emission line ratios, but it involves several assumptions and uncertainties. The location of the wind cannot be estimated from 1D spectra, and requires spatially-resolved spectroscopic observations.

While spatially-resolved spectroscopic observations offer more information regarding the outflow properties compared to 1D spectra, they are expensive to perform and the number of such observed sources is dramatically smaller than the number of sources with 1D spectra. The method suggested here can be used to estimate the mean location of the wind, and thus the mass outflow rate, in windy systems for which only 1D spectra are available. However, the method gives the mass-weighted (dust luminosity-weighted) centroid of the wind, while IFU-based studies usually define a range of distances. In some studies, the wind location is defined as the region beyond which broad emission lines are not detected (see e.g., \citealt{karouzos16a}). Therefore, the SED-based wind location is 2--3 times smaller than the definition of wind locations in such IFU-based studies.

To further examine our method and compare it to results that are based on spatially-resolved information, we use the sample presented by \citet{karouzos16a} and \citet{karouzos16b}. The sample consists of 6 type II AGN for which massive AGN-driven winds were detected through their 1D spectra \citep{woo16}, and were later observed with GMOS IFU on Gemini-North. We choose to compare to this sample since its properties and selection are similar to our work. Specifically, they study nearby AGN (redshifts of 0.05 to 0.1), they decompose the observed emission lines into narrow and broad components, and give robust identification of outflowing versus stationary gas components. Their wavelength range allowed them to measure dust-corrected [OIII] and H$\alpha$ luminosities in the outflowing component, where the dust extinction in the wind was estimated using the broad line components. Finally, their BPT classifications of the narrow and broad lines show that both the outflowing and the stationary gas are photoionized by the AGN.

Out of the 6 objects studied by \citet{karouzos16a}, we detect an additional greybody component in 4. We list in table \ref{t:comparsion_with_ifu} the dust temperatures of the additional greybody component, the AGN bolometric luminosity (equation \ref{eq:1}), and the estimated location of the wind (equation \ref{eq:2}). \citet{karouzos16a} estimated the size of the winds in their sample using two different methods. The first is the effective radius of the wind, which is based on either the broad [OIII] or H$\alpha$ emission line flux distribution. The second is the "kinematic" size of the wind, which is defined as the radius within which the broad [OIII] kinematics show evidence for non-gravitational motions. We list both of these estimates in table \ref{t:comparsion_with_ifu} for the [OIII] line. One can see that our SED-based estimate of the wind location is close to their effective wind radius. This is expected since our method results in a luminosity-weighted average location. On the other hand, the kinematic size, which is 2--3 times larger, includes all regions with a detected wind, even if the outer regions constitute a negligible fraction of the wind luminosity and ionized gas mass. Thus, while the "kinematic" size represents the maximal extent of the wind, our method gives its mass-weighted average location, which is consistent with IFU-based estimates. In a forthcoming paper, we estimate the mass outflow rates in a large sample of type II AGN for which only 1D spectra are available, taken from the parent sample we analysed in this work (Baron \& Netzer 2018). 

\begin{table*}
	\centering 
	\tablewidth{0.8\linewidth} 
\begin{tabular}{|l|l|l|l|l|l|}
ID &  $\mathrm{T_{dust}}$ [K] & $\mathrm{\log L(AGN)}$ [erg/sec] & $\mathrm{R_{dust}}$ [kpc] & $\mathrm{r_{eff}([OIII])}$ [kpc] & $\mathrm{r_{kin}([OIII])}$ [kpc]\\
 \hline
 \hline 
 J140453+532332 & 120 & 44.9 & 0.51 & 0.69 & 2.07 \\
 J160652+275539 & 120 & 44.3 & 0.15 & 0.43 & 1.29 \\
 J162233+395650 & 110 & 45.1 & 0.49 & 0.63 & 1.58 \\
 J172038+294112 & 110 & 45.8 & 0.34 & 0.64 & 1.92 \\
 \hline
\end{tabular}
\caption{Comparison of the wind location obtained through our NIR-MIR SED fitting to the IFU-based wind location of \citet{karouzos16a}. Col. 1: target ID, Col. 2: the best-fitting greybody temperature from the NIR-MIR SED fitting, Col. 3: the AGN bolometric luminosity, Col. 4: mass-weighted wind location according to the SED fitting method, Col. 5: the IFU-based effective radius of [OIII], Col. 6: the IFU-based kinematic size of the wind using the [OIII] emission line.}
\label{t:comparsion_with_ifu}
\end{table*}

Our method can provide an important additional information about IFU measurements of AGN-driven winds. Outflows are defined using emission line components that show evidence for non-gravitational motions. Thus, most detected outflows are those where the outflow direction is close to the line of sight, and outflows that are perpendicular to the line of sight are rarely detected (see example in \citealt{baron18}). In the former case, the velocity of the outflow is only slightly affected by projection effects, but the location and the size of the wind are difficult to determine. For example, in a conic-shape flow, the IFU-based size represents the opening angle of the outflow rather than its extent. Since the SED-based method is not sensitive to the inclination of the outflow with respect to the line of sight, it can be combined with the IFU-based size to constrain the inclination of the wind. 

IFU observations are also limited in their ability to resolve the wind, with the most significant systematic being the beam-smearing in ground-based, seeing-limited, observations. Bean smearing often results in a significant overestimation of the outflow extent, and thus the outflow energetics (see e.g., \citealt{husemann16, villar_martin16}). Indeed, HST observations reveal AGN-driven outflows which are more compact than those reported using ground-based IFU observations (e.g., \citealt{fischer18, tadhunter18}). The method presented here can be used to estimate the outflow extent in such cases. Unfortunately, the minimal dust temperature that can be detected depends on the redshift of the system, for example, in our case the central wavelength of the WISE W4 band. 

Winds in active galaxies can be multi-phased \citep{cicone18}, consisting of molecular, atomic, and ionized gas. It is unclear whether these phases are connected, and whether all three are simultaneously present in a given source. These phases can have different velocities, covering fractions, and outflow rates \citep{fiore17}. In particular, each gas phase is dusty, and would contribute to the total IR SED of the system, with a peak wavelength corresponding to the dust location and temperature, and the covering fraction of the specific component. Some of these phases are traced by emission lines and thus their location can be derived from IFU spectroscopy, for example, dusty ionized gas flows that are traced by optical emission lines (e.g., \citealt{karouzos16a, baron17b, rupke17, baron18}), or molecular outflows that are traced by sub-mm emission lines (e.g., \citealt{feruglio10, cicone14, sun14, veilleux17}).

Some outflowing phases are traced by absorption lines, and thus their location estimation strongly depends on the background source distribution. For example, neutral and molecular outflows are detected via optical (NaID) and FIR (OH) absorption lines \citep{rupke02, rupke05, rupke13, veilleux13}. In particular, the NaID absorption strength shows strong correlation with the dust reddening in AGN (e.g., \citealt{baron16}), thus the MIR emission of the dust in the outflow can be used to estimate the location of these neutral winds. Broad absorption line quasars (BALs) are systems that show strong outflows via UV absorption lines, and these too show correlation to the dust reddening along the line of sight (e.g., \citealt{reichard03}), which can be used to estimate the location of the flows. The estimation of the wind location in such cases does not follow exactly the methodology presented here, and requires additional modelling, which is beyond the scope of the present paper.

\subsection{Method evaluation and main uncertainties}\label{s:sed_fitting_uncertainties}

There are several sources of uncertainties associated with the NIR-MIR SED-fitting method. First, we used the Dn4000-based SFR measurement to select the corresponding SF template, which we then subtracted from the observed SEDs. In extreme cases, this method can be affected by scattered AGN light (e.g., \citealt{zakamska05}). In addition, for a given L(IR), there is a range of MIR SEDs that describe pure star forming galaxies. \citet{dominguez14} compared different SFR estimates for a sample of star forming and AGN-dominated galaxies. They found a tight correlation between the Dn4000-based and the FIR-based SFR measurements, with a scatter of 0.2--0.3 dex, some of which may be related to scattered AGN light. A difference of 0.3 dex in SFR estimation results in a factor of 2--3 difference in the luminosity of the chosen SF template at MIR wavelengths. This can affect the detection of the dusty component and in the determination of the dust temperature. Fortunately, as shown earlier, our fitting procedure does not require an additional dusty component in the majority of the non-windy systems, thus this is not a major source of uncertainty for most of the objects in our sample. Moreover, the uncertainty in the MIR SED due to the uncertain SFR is much smaller than the uncertainty due to torus normalization, as we argue below. Nevertheless, including FIR measurements can improve the quality of the SED fitting.

The torus normalisation is an additional source of uncertainty. The torus luminosity at 20 $\mathrm{\mu m}$ is estimated using the AGN bolometric luminosity, which in turn is estimated using narrow emission line luminosities. The scatter in the relation between narrow line luminosities and the AGN luminosity is roughly 0.4 dex \citep{netzer09}, and the scatter in the relation between the bolometric luminosity and MIR luminosity is about an order of magnitude \citep{netzer13, ichikawa17}. For this reason we allowed such a large factor (6) in the SED normalization. Since the dusty wind emits at the same wavelength range as the torus, the two are somewhat degenerate. Indeed, we have shown that our procedure misses hotter dust components  due to this degeneracy. A possible way to reduce this uncertainty is using X-ray selected AGN for which the MIR torus luminosity can also be estimated from the X-ray luminosity and absorbing column, perhaps with a smaller scatter \citep{lusso11, netzer13, stern15, chen17, ichikawa17, brown18}.

The exact determination of the dust temperature, which determines the derived location of the outflow, is hampered by the limited, broad band MIR data used in this study, in particular the fact that most of these temperatures were derived using only three MIR WISE bands, W2, W3 and W4. This results in an uncertainty of at least 20 K in our dust temperature estimation, which results in an uncertainty of 10\%--50\% in wind location estimation, and affects low dust temperatures more than high dust temperatures. A possible way to reduce this uncertainty is using MIR spectroscopy, e.g. $\mathit{Spitzer}$'s Infrared Spectrograph, where the temperature of the greybody component can be estimated with a smaller uncertainty (e.g., \citealt{mor12, lambrides18}). Furthermore, the lowest dust temperature we can detect using the SED-based method depends on the longest wavelength we use, thus we are not sensitive to colder dust components with larger wind extents. The wavelength coverage can be improved by adding $\mathit{Spitzer}$ spectroscopy and $\mathit{Herschel}$/PACS photometry although the number of observed sources is significantly smaller.

Our method was adopted for ionized gas outflows and its results are more uncertain in those cases where dust in molecular and atomic gas flows contribute, significantly, to the MIR emission. Specifically, the MIR SED will be dominated by the most-massive dusty gas component, or can include contributions from several such outflowing phases. We find that for a small number of objects in our sample, the quality of the fit improves when adding an additional greybody component to the model. However, due to the limited measurements we work with, we choose not to include such a component, even when it appears necessary. Given a better wavelength resolution and coverage, the SED of a system that hosts multi-phased outflows can be fitted with several greybody components.

\section{Summary and conclusions}\label{s:concs}

We use a sample of $\sim$4\,000 type II AGN at redshifts $0.05 \leq z \leq 0.15$ to study the infrared properties of AGN-driven winds. We divided the sample into two groups of roughly equal size, one in which ionized gas outflows are detected through optical emission lines and the second in which winds are not detected. We produced NIR-MIR stacked SEDs of the two samples by dividing them into bins of similar stellar mass and SFR, while controlling for the dust-corrected [OIII] luminosity. We found MIR emission excess in the stacked SEDs of the windy systems, compared to their non-windy control sample. Since the stellar mass and SFR are similar in the two samples, the excess cannot be due to the direct starlight contribution to the optical-NIR or SF in the galaxy. The similar [OIII] luminosity ensures that the torus luminosities in the two samples are also the same. We suggest that the MIR excess is due to dust which is mixed with the ionized gas in the outflow. This dust is heated by the AGN and emits at MIR wavelengths. We argue that such an emission component is unavoidable since the outflowing gas is dusty. Given the typical observed luminosities of the optical lines associated with the galactic-scale outflows, the dust emission signature has to be detected at MIR wavelengths. 

We fitted the SEDs of all the objects in our sample. The model SED includes contributions from a direct starlight component, torus emission, SF contribution, and a new greybody component that represents the dust in the outflow. We found that 80\% of the non-windy systems do not require an additional greybody component, while 86\% of the windy systems require an additional greybody component for a proper fit. We estimated the torus covering factor in the windy and non-windy systems, and find consistent covering factors in the two groups, which are also in line with torus covering factors in type I AGN. Thus, our SED fitting procedure successfully recovers the dusty component in most cases. 

The new MIR dust emission component can be used to put important new constraints on outflow properties, such as its location and covering factor. We used the best-fitting SEDs to estimate the mass-weighted average location of the wind for $\sim$1\,700 AGN. The distribution in location shows a prominent peak around 500 pc, with a tail extending to large distances ($\sim$10 kpc), and no systems with winds below 50 pc. We estimated the covering factor of this dusty component, and found a wide distribution that is centered around 10\%, with 24\% of the winds having covering factors larger than 20\%, and 8\% of the systems showing covering factors larger than 50\%.

So far, outflow locations could only be estimated using spatially-resolved spectroscopy. Since IFU observations are very expensive, the samples of windy systems for which the outflow location and mass outflow rate were estimated using such observations are small. Furthermore, IFU observations are subject to systematics such as projection effects and beam smearing, which can significantly affect the estimated outflow locations and energetics. The dust emission method is not sensitive to the various systematics affecting IFU observations, and can reduce the uncertainties involved in such measurements. While the method presented in this work cannot provide accurate measurements of outflow locations in individual systems, it can be used in combination with 1D spectra to estimate statistically the location and energetics of AGN-driven winds. This can be used to estimate mass outflow rates in thousands of active galaxies and is the subject of a forthcoming publication.

\section*{Acknowledgments}
We thank our referee for useful comments and suggestions that helped improve this manuscript. We thank R. Davies, N. Lubelchick, D. Lutz, J. X. Prochaska, I. Reis, and B. Trakhtenbrot for useful discussions regarding the manuscript.
This work is supported in part by the Israel Science Foundation grant 284/13.
The spectroscopic analysis was made using IPython \citep{perez07}. We also used the Python package astropy\footnote{www.astropy.org/}.

This work made use of SDSS-III\footnote{www.sdss3.org} data. Funding for SDSS-III has been provided by the Alfred P. Sloan Foundation, the Participating Institutions, the National Science Foundation, and the U.S. Department of Energy Office of Science. SDSS-III is managed by the Astrophysical Research Consortium for the Participating Institutions of the SDSS-III Collaboration including the University of Arizona, the Brazilian Participation Group, Brookhaven National Laboratory, Carnegie Mellon University, University of Florida, the French Participation Group, the German Participation Group, Harvard University, the Instituto de Astrofisica de Canarias, the Michigan State/Notre Dame/JINA Participation Group, Johns Hopkins University, Lawrence Berkeley National Laboratory, Max Planck Institute for Astrophysics, Max Planck Institute for Extraterrestrial Physics, New Mexico State University, New York University, Ohio State University, Pennsylvania State University, University of Portsmouth, Princeton University, the Spanish Participation Group, University of Tokyo, University of Utah, Vanderbilt University, University of Virginia, University of Washington, and Yale University. 

\bibliographystyle{mn2e}
\bibliography{ref_ir_properties}

\clearpage

\appendix

\section{Stacked SEDs: bins}\label{a:stacks_bins}

\begin{figure*}
\includegraphics[width=0.9\textwidth]{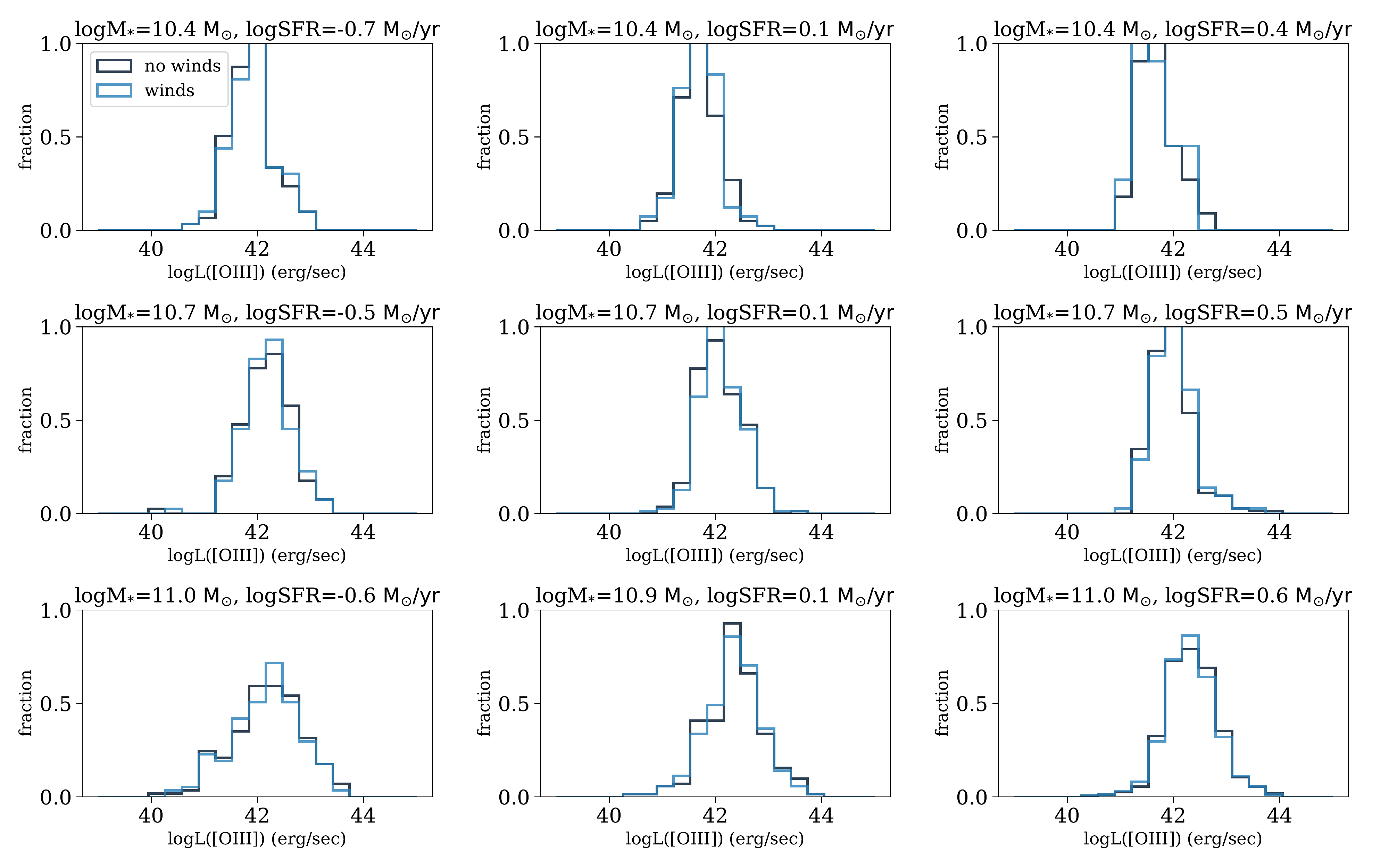}
\caption{The distribution of the dust-corrected [OIII] luminosity for systems that show ionized winds and systems that do not. Each panel resprests the distribution for different SFR and stellar mass bins. The objects within each bin are combined to a single stacked spectrum.}\label{f:oiii_histogram_of_bins}
\end{figure*}

\begin{figure*}
\includegraphics[width=0.9\textwidth]{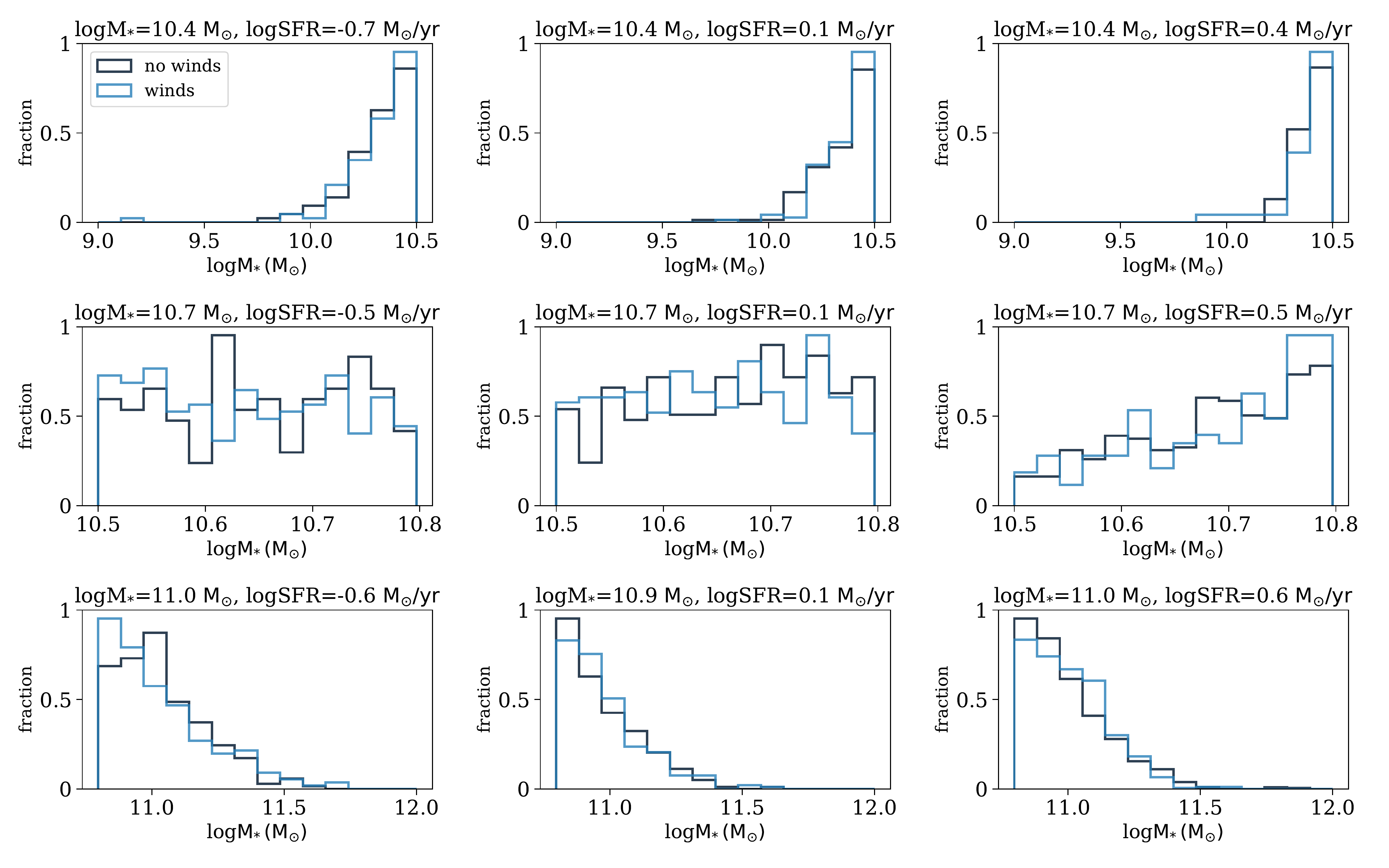}
\caption{The distribution of the stellar mass for systems that show ionized winds and systems that do not. Each panel resprests the distribution for different SFR and stellar mass bins. The objects within each bin are combined to a single stacked spectrum.}\label{f:mass_histogram_of_bins}
\end{figure*}

\begin{figure*}
\includegraphics[width=0.9\textwidth]{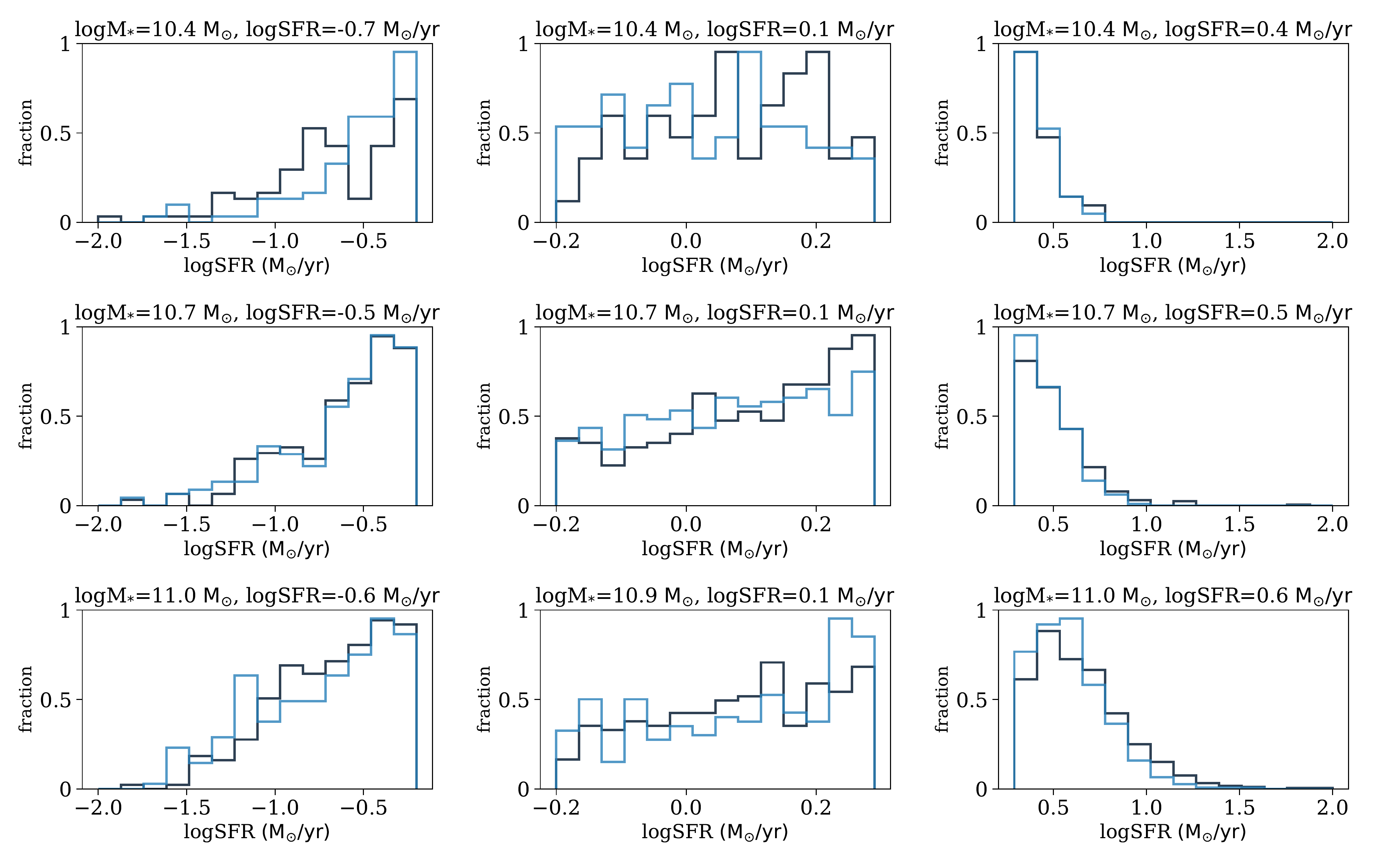}
\caption{The distribution of the SFR for systems that show ionized winds and systems that do not. Each panel resprests the distribution for different SFR and stellar mass bins. The objects within each bin are combined to a single stacked spectrum.}\label{f:sfr_histogram_of_bins}
\end{figure*}

\begin{table*}
	\centering 
	\tablewidth{0.8\linewidth} 
\begin{tabular}{||c|l|l|l||}
 SFR / stellar mass bins &  $\mathrm{9.0\, M_{\odot} \,\, to\,\,10.5\, M_{\odot}}$ &  $\mathrm{10.5\, M_{\odot} \,\, to\,\,10.8\, M_{\odot}}$ & $\mathrm{10.8\, M_{\odot} \,\, to\,12 \,\, M_{\odot}}$ \\
 \hline
 \hline 
 $\mathrm{-2.0\,\, to\, -0.2\, M_{\odot}\,yr^{-1}}$ & 94 &  126 & 181 \\
 $\mathrm{-0.2\,\, to\,\, 0.3\, M_{\odot}\,yr^{-1}}$ &  129 &  253 & 225 \\
 $\mathrm{0.3\,\, to\,\, 2.0\, M_{\odot}\,yr^{-1}}$ &  35 &  229 & 513 \\
 \hline
\end{tabular}
\caption{We divide the objects into bins of stellar mass and SFR. For each bin, we select windy and non-windy systems to have similar distribution in [OIII] luminosity. The table gives the bins we used, and the number of objects per bin. The number of windy and non-windy systems per bin is similar, thus the total number of objects in each bin is twice the value given.}
\label{t:bins}
\end{table*}

\end{document}